\documentclass[aps,prb,reprint,superscriptaddress]{revtex4-2}
\usepackage{amsmath,amssymb}
\usepackage{graphicx}% Include figure files
\usepackage{dcolumn}% Align table columns on decimal point
\usepackage{bm}% bold math
\usepackage{booktabs}
\usepackage{hyperref}% add hypertext capabilities
\begin{document}

\title{Symmetry classes of hyperdeterminant wavefunctions: Application to phase transitions between a bosonic fractional Chern insulator and superfluids}
\author{Guan-Lin Lin}
\affiliation{Department of Physics, Boston College, Chestnut Hill, MA, 02467, USA}
\author{Xiaodong Hu}
\affiliation{Department of Materials Science and Engineering, University of Washington, Seattle, WA 98195, USA}
\author{Di Xiao}
\affiliation{Department of Materials Science and Engineering, University of Washington, Seattle, WA 98195, USA}
\affiliation{Department of Physics, University of Washington, Seattle, WA 98195, USA}
\author{Ying Ran}
\affiliation{Department of Physics, Boston College, Chestnut Hill, MA, 02467, USA}
\begin{abstract}
We study a phase transition between a superfluid phase and a fractional Chern insulator phase in a hard-core boson model on the honeycomb lattice using symmetric hyperdeterminant (Hdet) wavefunctions. Combining projective symmetry group analysis with variational Monte Carlo, we construct and optimize Hdet states in various symmetry classes. Based on the associated effective theories, we show that different symmetry classes, i.e., representations of the projective symmetry group, describe specific quantum phases. Compared with traditional fermionic parton construction, the symmetric Hdet wavefunctions are shown to improve microscopics significantly, allowing quantitative investigation of quantum phase diagrams. We argue that, in contrast with the common belief, this putative continuous phase transition is described by a fluctuating $SU(2)_g$ gauge field coupling with four Dirac nodes forming two $SU(2)_g$ doublets (i.e., $N_f=2$ QCD). 
\end{abstract}
\maketitle
\section{Introduction}\label{sec:intro}
Conventional phase transitions are described by Landau-Ginzburg theory in terms of symmetry breaking. When fractionalized degrees of freedom (dof) are present in quantum many-body systems, exotic quantum phase transitions beyond this traditional paradigm can happen and have been explored \cite{Senthil2004,SenthilVojtaSachdev2004,Senthil2008,Isakov2012,Nandkishore2012,SongZhang2023}. As a particular example, field-theory studies have proposed the possibility of a continuous phase transition between a superfluid (SF) and a fractional Chern insulator (FCI) \cite{Barkeshli2014,BarkeshliYaoLaumann2015,SongZhangSenthil2024}. Recent numerical simulations found strong evidence for such a transition in bosonic models \cite{Lu2025,Wang2026}. In fermionic systems, it has been argued that a related exotic continuous phase transition from a Fermi liquid to a composite Fermi liquid can occur \cite{Barkeshli2012}. Given the exciting experimental discovery of the fractional quantum anomalous Hall (FQAH) states \cite{Cai2023,Zeng2023,Park2023,Xu2023,Lu2024,Xie2025}, the nature of these exotic phase transitions warrants further investigation, as they may be realized experimentally in the foreseeable future.

Fractional Chern insulators can arise from interacting particles in topological lattice bands \cite{Haldane1988,Wang2011,Tang2011,Sun2011,Neupert2011,Sheng2011,RegnaultBernevig2011,Xiao2011}.
In these fractionalized phases, the low-energy physics can be described in terms of anyonic quasiparticles coupling with emergent gauge fields. In order to investigate the nature of such exotic phase transitions in microscopic models, numerical simulations are usually unavoidable and can be computationally demanding. Ideally, one would like reliable simulation techniques that can directly access those fractionalized dof. For instance, one would like to compute the microscopic band structure of fractionalized quasiparticles. This poses one of the major challenges from a theoretical perspective: most existing numerical techniques have direct access only to the dof of gauge-neutral physical particles. Extracting information about fractionalized quasiparticles from physical correlation functions and numerical calculations is indirect and often nontrivial \cite{Essin2014,He2017,Wang2026}. 

On the other hand, parton constructions provide a microscopic description in terms of fractionalized degrees of freedom \cite{Baskaran1987,AffleckMarston1988,Jain1989,Wen1991,LeeNagaosaWen2006}. Together with variational Monte Carlo (VMC) simulations, parton constructions have been shown to be very useful techniques in the context of quantum spin liquids (QSL) and FCI \cite{Ran2007,McGreevy2012,Lotric2026}. However, as we will see shortly, the traditional parton constructions have limitations that will be unable to universally capture microscopic details of the wavefunctions in a quantum phase diagram. 

Consider a lattice model with one hard-core boson orbital $|\mathbf r^{(b)}\rangle$ per site $\mathbf r$, whose second-quantized operator is $b_{\mathbf r}$. The parton construction enlarges the physical Hilbert space via introducing auxiliary, fermionic dof. The traditional parton construction can be understood via the following fusion operator:
\begin{align}
\hat{\mathbf F}^{\rm trad.}_{\mathbf r}\equiv b^\dagger_{\mathbf r}f^{(1)}_{\mathbf r}f^{(2)}_{\mathbf r},\label{eq:traditional_parton_construction}
\end{align}
where $f^{(p)}_{\mathbf r}$ are the fermionic parton operators of two species ($p=1,2$). Throughout this paper, \emph{we denote the parton species as the parton color, or simply color}. This fusion operator is a linear map between the local Fock space of the partons and that of the physical bosons. The physical bosonic many-body wavefunction can be obtained as:
\begin{align}
|\Psi^{(b)}\rangle=&\hat {\mathbf F}^{N_b}|\Psi^{MF}\rangle,&\hat {\mathbf F}\equiv& \sum_{\mathbf r}\hat{\mathbf F}_{\mathbf r}\label{eq:parton_fused_into_physical}
\end{align}
where $|\Psi^{MF}\rangle\equiv |\Psi^{(1)}\rangle\otimes |\Psi^{(2)}\rangle$ is the many-body parton mean-field state: a tensor product of two Slater determinants $|\Psi^{(p)}\rangle$, one for each parton color. $|\Psi^{(p)}\rangle$ is specified by the \emph{filled single-particle orbitals}. Since the number of partons in each color should be equal to that of the physical boson $N_b$, the wavefunctions of these $N_b$ linearly independent filled orbitals can be labeled as $\phi_{\mathbf r}^{(p),i}$ ($i=1,...,N_b$). The partons are exactly the fractionalized dof, which is why this technique can access them by construction, providing mean-field-like, intuitive physical pictures.

In recent works \cite{Hu2024,Lin2026,Hu2026}, we generalized the traditional fermionic parton constructions into hyperdeterminant (Hdet) wavefunctions. In the present case, \emph{an Hdet state with a locality structure} can be understood by a generalized fusion operator:
\begin{align}
\hat{\mathbf F}^{\rm Hdet}_{\mathbf r}\equiv b^\dagger_{\mathbf r}\sum_{\alpha=1}^{R}\lambda_{\mathbf r,\alpha} f^{(1)}_{\mathbf r,\alpha}f^{(2)}_{\mathbf r,\alpha},\label{eq:Hdet_parton_construction}
\end{align}
where on each site $\mathbf r$, we introduce $R$ parton orbitals $f^{(p)}_{\mathbf r,\alpha}$, $\alpha=1,...,R$ for each color-$(p)$. Each orbital $\alpha$ labels a fusion channel, and $\lambda_{\mathbf r,\alpha}$ is the corresponding fusion amplitude. The many-body wavefunction $|\Psi^{(b)}\rangle$ of physical bosons is still obtained as Eq.~(\ref{eq:parton_fused_into_physical}), which can be shown to be a hyperdeterminant wavefunction of a rank-3 fusion tensor $T_{\mathbf r,j,k}$:
\begin{align}
T_{\mathbf r,j,k}\equiv \langle \mathbf r^{(b)}|\hat{\mathbf F}|\phi^{(1),j}\rangle|\phi^{(2),k}\rangle=\sum_{\alpha=1}^R\lambda_{\mathbf r,\alpha}\phi^{(1),j}_{\mathbf r\alpha}\phi^{(2),k}_{\mathbf r\alpha},\label{eq:CPD}
\end{align}
where $j,k=1,...,N_b$, while $\mathbf r$ runs through the lattice. Precisely speaking, $|\Psi^{(b)}\rangle$ is defined via the overlap with the real-space occupation basis (see Ref.~\cite{Lin2026} for details):
\begin{align}
\langle \mathbf r_{i_1},\mathbf r_{i_2},...,\mathbf r_{i_{N_b}}|\Psi^{(b)}\rangle\equiv \text{Hdet}(T^{\rm sub}),\label{eq:Hdet_def}
\end{align}
where $\rm Hdet$ is the combinatorial hyperdeterminant of the cubic tensor $T^{\rm sub}$, which is obtained by selecting $\mathbf r_{i_1},\mathbf r_{i_2},...,\mathbf r_{i_{N_b}}$ slices of the full fusion tensor $T_{\mathbf r,j,k}$.

If one fixes $\mathbf r$, Eq.~(\ref{eq:CPD}) can be understood as the rank factorization of the square matrix $T_{jk}[\mathbf r]\equiv T_{\mathbf r,j,k}$, and $R$ is its rank. The traditional construction $\hat{\mathbf F}^{\rm trad.}_\mathbf r$ corresponds to $R=1$. As shown in Ref.~\cite{Lin2026}, there are seminal wavefunctions that necessarily require $R>1$. For instance, the $\nu=\frac{2}{3}$ Jain's fractional quantum Hall wavefunction for bosons has $R=3$.

Allowing $R>1$, the key new ingredient in the Hdet construction, is important for encoding the microscopic details of physical many-body wavefunctions. As the \textbf{first main result} of this paper, we will show that \emph{Hdet states with $R>1$ universally capture the quantum phase diagram, providing reliable energetics that the traditional $R=1$ construction cannot achieve.}

WLOG, it is always possible to absorb the fusion amplitudes into the wavefunction of filled parton orbitals, e.g., $\lambda_{\mathbf r,\alpha}\phi^{(1),j}_{\mathbf r\alpha}\equiv \tilde\phi^{(1),j}_{\mathbf r\alpha}$, after which all the fusion amplitudes are equal to unity:
\begin{align}
\text{Canonical choice: }\lambda_{\mathbf r,\alpha}=1,\forall\mathbf r,\alpha.
\end{align}
To simplify the discussion, \emph{we will work within this canonical choice from now on}:
\begin{align}
&\hat{\mathbf F}^{\rm Hdet}_{\mathbf r}\equiv b^\dagger_{\mathbf r}P_\mathbf r,\text{ where } P_\mathbf r\equiv\sum_{\alpha=1}^{R} f^{(1)}_{\mathbf r,\alpha}f^{(2)}_{\mathbf r,\alpha}=\frac{1}{2}\psi^T_\mathbf r J \psi_\mathbf r,\notag\\
&J\equiv\begin{pmatrix} 0&I_R\\-I_R&0\end{pmatrix},\;\psi_\mathbf r\equiv (f^{(1)}_{\mathbf r,1},...,f^{(1)}_{\mathbf r,R},f^{(2)}_{\mathbf r,1},...,f^{(2)}_{\mathbf r,R})^T.\label{eq:Hdet_canonical_parton_construction}
\end{align}

As in all parton constructions, Eq.~(\ref{eq:Hdet_canonical_parton_construction}) enlarges the physical Hilbert space, leading to gauge redundancies characterized by a gauge group GG. GG is defined to be the gauge group such that $\forall W(\mathbf r)\in GG$, $W(\mathbf r)|\Psi^{MF}\rangle$ and $|\Psi^{MF}\rangle$ give identical physical wavefunction $|\Psi^{(b)}\rangle$ after fusion. Precisely speaking, GG$=USp(2R)$ (when $R=1$, GG$=USp(2)\cong SU(2)$) is defined as the subgroup of the unitary rotations among the $2R$ parton-orbitals: $\psi^{\dagger T}_{\mathbf r}\rightarrow W(\mathbf r)\psi^{\dagger T}_{\mathbf r}$ (or equivalently $\psi_{\mathbf r}\rightarrow W^*(\mathbf r)\psi_{\mathbf r}$), leaving the fusion operator $\hat{\mathbf F}^{\rm Hdet}_{\mathbf r}$ invariant:
\begin{align}
\text{GG}=USp(2R)=\{W(\mathbf r)\in U(2R)|W(\mathbf r)^T J W(\mathbf r)=J\}.
\end{align}

For example, a specific subgroup $SU(2)_g\subset GG$ represents the gauge redundancy in the color space:
\begin{align}
\forall W(\mathbf r)\in SU(2)_g,\;\;f^{(p)\dagger}_{\mathbf r,\alpha}\rightarrow \sum_{p'}W_{p,p'}(\mathbf r)f^{(p')\dagger}_{\mathbf r,\alpha}.
\end{align}
As pointed out by Wen \cite{Wen2002} (see Ref.~\cite{Lin2026} for details of effective theories of Hdet states in terms of the variational manifold path integral), since the parton mean-field state $|\Psi^{MF}\rangle$ at the saddle point may reduce the GG gauge redundancy to a subgroup: the invariant gauge group (IGG) via the Anderson-Higgs mechanism, the gauge fluctuations in the low-energy effective theory is really determined by IGG. For example, if one can choose a gauge such that $|\Psi^{(1)}\rangle$ and $|\Psi^{(2)}\rangle$ are identical at the saddle point (i.e., color permutation symmetric), then a \emph{site-independent} $SU(2)_g$ gauge transformation leaves $|\Psi^{MF}\rangle$ invariant. If these are the only gauge transformations leaving $|\Psi^{MF}\rangle$ invariant, then IGG$=SU(2)_g$, and the low-energy effective theory is fermionic partons coupling with the dynamical $SU(2)_g$ gauge field. If such a gauge choice is impossible when the $SU(2)_g$ gauge fluxes in $|\Psi^{MF}\rangle$ Higgs out the $SU(2)_g$ down to $U(1)_g$:
\begin{align}
U(1)_g\equiv \{W(\mathbf r)=e^{i\theta_{\mathbf r}\tau^z}\}\subset SU(2)_g,\label{eq:U1_g_def}
\end{align}
the low-energy effective theory is fermionic partons coupling to the dynamical $U(1)_g$ gauge field. ($\tau^z$ is the Pauli matrix in the color space.)

In fact, the original theoretical proposal \cite{Barkeshli2014} of the FCI-SF continuous phase transition was exactly based on the traditional parton construction Eq.~(\ref{eq:traditional_parton_construction}), \emph{under an important assumption}. The intuitive picture for the proposed transition is a band inversion of the filled parton states: In the FCI phase, parton-$(1)$ (parton-$(2)$) fills a gapped Chern band with Chern number $\mathcal C^{(1)}=1$ ($\mathcal C^{(2)}=1$). In the SF phase, parton-$(1)$ still fills a gapped band with $\mathcal C^{(1)}=1$, but parton-$(2)$ fills a gapped band with $\mathcal C^{(2)}=-1$. \emph{The underlying assumption is that the IGG$=U(1)_g$ for both phases}, at least in the neighborhood of the phase transition. One key question this paper addresses is exactly this underlying assumption. As the \textbf{second main result}, we find that, at least in the model we investigate, \emph{IGG$=SU(2)_g$ on the FCI side, while IGG$=U(1)_g$ on the SF side.}

\begin{figure}
    \centering
    \includegraphics[width=0.9\linewidth]{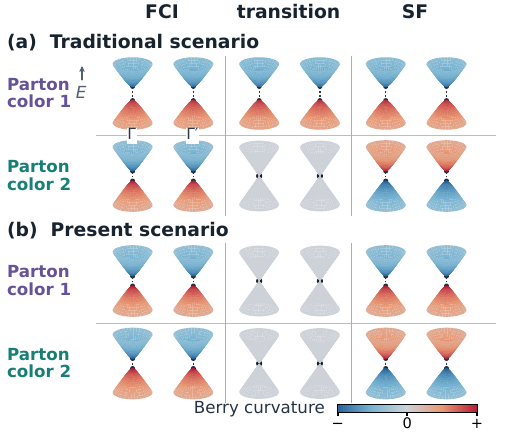}
    \caption{The illustration for the traditional scenario (a) and the scenario found in the present work (b) for the putative FCI to SF continuous phase transition in the parton picture, where band inversion of parton color-(2) occurs at the two magnetic-transition-related momentum valleys, $\Gamma$ and $\Gamma'$, leading to a change of the Chern number by 2. In the scenario found in the present study, the parton color permutation symmetry is respected inside the FCI phase and at the transition point; consequently, the parton band gaps for both colors must close simultaneously at the transition.}
    \label{fig:two_scenarios}
\end{figure}

If IGG$=U(1)_g$ on both sides of the phase transition, the band-inversion intuitive picture would be simple: the band gap of parton-$(1)$ remains finite throughout the phase transition, and only parton-$(2)$ experiences band-gap closing and reopening via two Dirac nodes, changing the Chern number by 2 for its filled band (See Fig.~\ref{fig:two_scenarios}(a) for an illustration). This is the commonly believed scenario for such a phase transition. If one integrates out the parton matter fields, the $U(1)_g$ gauge field has a level-2 Chern-Simons term on the FCI side, capturing its topological order. On the SF side, the Chern-Simons term cancels out, and the gauge dynamics is described by Maxwell photons. The linearly dispersing photon mode corresponds to the Goldstone mode of the SF phase under boson-vortex duality \cite{FisherLee1989}. The phase transition is described by $N_f=2$ $QED_3$: two gapless Dirac nodes from color-$(2)$ coupling with the dynamical $U(1)_g$ gauge field.

This is different from what we found in the present model study. Based on both numerical evidence and a sharp projective symmetry argument, after choosing the proper gauge, the two parton-mean-field states need to satisfy $|\Psi^{(1)}\rangle=|\Psi^{(2)}\rangle$ inside the FCI phase, and therefore IGG$=SU(2)_g$. The Higgs term reducing $SU(2)_g$ down to $U(1)_g$ is forbidden by the $C_2$ lattice rotation symmetry inside the FCI phase due to the 2-fold ground state degeneracy (GSD) on a torus. Therefore, the Higgs transition, reducing $SU(2)_g$ down to $U(1)_g$, and the parton band inversion have to occur at the same point: the FCI-SF transition point. In this scenario, the band inversion is described by a simultaneous gap closing and reopening of parton-$(1)$ and parton-$(2)$, each having two Dirac nodes. The transition is described by $N_f=2$ $QCD_3$: four gapless Dirac nodes from both parton-$(1)$ and parton-$(2)$, forming two $SU(2)_g$ doublet coupling with the dynamical $SU(2)_g$ gauge field. (See Fig.~\ref{fig:two_scenarios}(b) for an illustration.) Inside the FCI phase, after integrating out the parton matter field, one obtains the level-1 $SU(2)_g$ Chern-Simons theory, also capturing the topological order.

\section{Model and projective-symmetry analysis}\label{sec:psg}
We study a hard-core-boson tight-binding model on the honeycomb lattice up to third-nearest-neighbor hopping. In the absence of interactions ($V_1=V_2=0$), a robust FCI ground state was found when the lower Chern band is half-filled ($t=1,\ t^\prime =0.6,\ t^{\prime\prime}=-0.58,\ \phi=0.4\pi$) \cite{Wang2011}. As the Chern band gradually goes dispersive by tuning only $t^{\prime \prime}$, phase transitions from FCI to two different SF phases: SF(M) and SF($\Gamma$) were reported in previous simulations \cite{Lu2025,Lotric2026} based on density matrix renormalization group \cite{White1992} and variational Monte Carlo.

\begin{align}
    &H=-\sum_{\langle \mathbf r,\mathbf r' \rangle}t(b_\mathbf r^\dagger b_{\mathbf r'}+\text{h.c.})-\sum_{\langle \langle \mathbf r,\mathbf r' \rangle \rangle}t^\prime(e^{i\phi}b_\mathbf r^\dagger b_{\mathbf r'}+\text{h.c.})\nonumber\\
    &-\sum_{\langle \langle \langle \mathbf r,\mathbf r' \rangle \rangle \rangle}t^{\prime \prime}(b_\mathbf r^\dagger b_{\mathbf r'}+\text{h.c.})+V_1\sum_{\langle \mathbf r,\mathbf r' \rangle} n_\mathbf r n_{\mathbf r'}+V_2\sum_{\langle \langle \mathbf r,\mathbf r' \rangle \rangle} n_\mathbf r n_{\mathbf r'}\label{eq:hard_core_boson_model}
\end{align}
\begin{figure}[t]
    \centering
    \includegraphics[width=0.46\columnwidth]{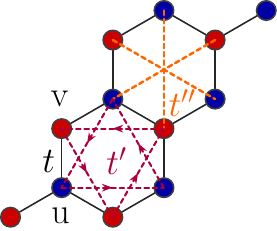}
    \hfill
    \includegraphics[width=0.5\columnwidth]{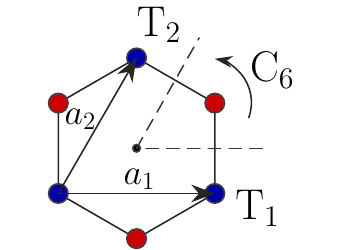}
    \caption{Honeycomb hard-core boson model, up to third nearest hoppings (left) and symmetry operations on the honeycomb lattice (right).}
    \label{fig:honeycomb_model_symmetry}
\end{figure}

SF($\Gamma$) and SF($M$) differ by the momentum points of the boson condensation: the $\Gamma$-point and the three $M$-points, respectively. Consequently, SF($\Gamma$) does not break lattice symmetries, leading to a single ground state on a finite-size sample; while SF($M$) does break lattice symmetries, giving multiple ground states. The continuous phase transition between SF($\Gamma$) and FCI was reported to occur at $t''=-0.495$, while a first-order phase transition between FCI and SF($M$) was reported to happen at  $t''\sim -0.7$. 

In a very recent work \cite{Hu2026}, we have applied the general Hdet variational states without a locality structure to study this model, where excellent energetics and overlaps are found. In the present work, we instead focus on Hdet states with the locality structure since it has clear associated effective quantum field theories \cite{Lin2026}. Namely, with the locality structure and a fixed $R$, we can construct variational Hdet states for each quantum phase under investigation separately, allowing us to extract rich underlying physics of the fractionalized degrees of freedom.

In particular, to understand the continuous transition between FCI and SF($\Gamma$) at the microscopic level, we begin by constructing the Hdet wavefunctions for these two phases, \emph{both of which respect lattice symmetries}. Mathematically, we need to identify both the projective symmetry group (PSG) and the \emph{representation} of PSG realized by the parton mean-field state $|\Psi^{MF}\rangle$.

\textbf{PSG preserving color: } As shown in the seminal work by Wen \cite{Wen2002}, the parton only needs to respect symmetry in a projective fashion, leading to the algebraic understanding of symmetry fractionalization phenomena. Here, the full symmetry group $SG$ of the model Eq.~(\ref{eq:hard_core_boson_model}) is generated by $U(1)_c$-charge conservation, lattice translation $T_1,T_2$, $C_6$ rotation (see Fig.~\ref{fig:honeycomb_model_symmetry} for an illustration):
\begin{align}
    T_1:(x,y,w) &\rightarrow (x+1,y,w),\ w=u,v\notag\\
    T_2:(x,y,w) &\rightarrow (x,y+1,w),\ w=u,v\notag\\
    C_6:(x,y,u) &\rightarrow (-y+1,x+y-1,v),\notag\\
    (x,y,v) &\rightarrow (-y,x+y,u).
    \end{align}
where $w=u,v$ label the two sublattices and a generic site on the honeycomb lattice is represented as $\mathbf r=(x,y,w)$. These $SG$ generators on a finite torus with $N\times N$ unit cells ($N$ is an even integer to ensure that the half-filling is well-defined) satisfy group identities:
\begin{align}
    T_1^{-1}T_2T_1T_2^{-1}&=\mathbf e  ,\notag\\ 
    T_1^{-1}C_6T_1T_2^{-1}C_6^{-1}&=\mathbf e,\notag\\ 
    T_2^{-1}C_6T_1C_6^{-1}&=\mathbf e,\notag\\
    C_6^6&=\mathbf e,\notag\\
    T_1^N=T_2^N&=\mathbf e,\label{eq:sg_identities}
\end{align}
where the last identity is due to the periodic boundary condition.

The assumption of the projective symmetry group (PSG) is that a symmetry operation $g\in SG$ is implemented as:
\begin{align}
U_g\equiv W_g \cdot g,
\end{align}
where $W_g\in GG$ is a gauge transformation associated with $g$. Since the physical boson is half-filled, in order for each color to have a gapped band insulator state, the translations $T_1,T_2$ need to be implemented projectively on both color as $\pi$-flux states: 
\begin{align}
U_{T_1}U_{T_2}=(-1)U_{T_2} U_{T_1}.\label{eq:pi_flux_condition}
\end{align}

We begin with the PSG analysis under the assumption that $W_g=e^{i\phi_{g}(\mathbf r,\alpha)\tau^z}$ is a gauge transformation corresponding to a site- and orbital-dependent phase factor:
\begin{align}
W_g: f^{(1)\dagger}_{\mathbf r,\alpha}\rightarrow e^{i\phi_{g}(\alpha,\mathbf r)} f^{(1)\dagger}_{\mathbf r,\alpha},\;\; f^{(2)\dagger}_{\mathbf r,\alpha}\rightarrow e^{-i\phi_{g}(\alpha,\mathbf r)} f^{(2)\dagger}_{\mathbf r,\alpha}.
\end{align}
This assumption is motivated by the fact that when $R=1$, GG$=SU(2)_g$, and such an $U(1)$ phase factor is the most general GG transformation without hybridizing the two colors, which is the natural way to implement the symmetry projectively. In fact, as we will see soon, this assumption is not always satisfied by the Hdet saddle-points in the FCI phase -- one of the key findings of the present work. 

As shown in Appendix~\ref{app:psg_derivation}, after fixing the gauge, the algebraic equations satisfied by PSG elements $\{U_g\}$, i.e., projective generalizations of Eq.~(\ref{eq:sg_identities}), can be fully fixed:
\begin{subequations}\label{eq:psg_group_identities}
\begin{align}
U_{T_1}U_{T_2}U_{T_1}^{-1}U_{T_2}^{-1}
	&=-1, \label{eq:psg_translation}\\
U_{T_1}^{-1}U_{C_6}U_{T_1}U_{T_2}^{-1}U_{C_6}^{-1}
	&=1, \label{eq:psg_rotation_1}\\
U_{T_2}^{-1}U_{C_6}U_{T_1}U_{C_6}^{-1}
	&=1, \label{eq:psg_rotation_2}\\
U_{C_6}^{6}
	&=1, \label{eq:psg_rotation_6}\\
U_{T_1}^{N}
	&=e^{i\varphi_1\tau_z}, \label{eq:psg_boundary_1}\\
U_{T_2}^{N}
	&=e^{i\varphi_2\tau_z}.
	\label{eq:psg_boundary_2}
\end{align}
\end{subequations}
After fixing the gauge, all $W_g=e^{i\phi_{g}(\mathbf r,\alpha)\tau_z}$ gauge transformations can be determined:
\begin{align}
    \phi_{T_1}(x,y,w,\alpha)&= \frac{\varphi_1}{N},\notag\\
    \phi_{T_2}(x,y,w,\alpha)&=x\pi+\frac{\varphi_2}{N},\notag\\
    \phi_{C_6}(x,y,w,\alpha)&=-\frac{\pi}{2}-\frac{\varphi_2}{2N}+[xy+\frac{(x-1)x}{2}]\pi\notag\\
    &+\frac{\pi l_\alpha}{3}+\frac{1}{N}[(\varphi_2-\varphi_1)y+\varphi_2x].\label{eq:psg-phase}
\end{align}
Here, $l_\alpha=0,1,2$ ($\alpha=1,...,R$) can be interpreted as the three possible site-centered $C_3$ rotation angular momentum carried by the parton orbital $\alpha$, and should be treated as discrete variational parameters of Hdet wavefunctions. In our numerical simulations up to $R=3$, we find that choosing $l_\alpha=0$ $\forall \alpha$ gives Hdet states that are lowest in energy throughout the investigated regime. Consequently, we fix $l_\alpha=0$ $\forall \alpha$ below.

Before proceeding, we define a few useful subgroups of the PSG:
\begin{align}
C_6\text{-PSG}\equiv & \text{group generated by } \{U_{T_1},U_{T_2},U_{C_6}\},\notag\\
C_3\text{-PSG}\equiv& \text{group generated by } \{U_{T_1},U_{T_2},U_{C_3}\equiv U_{C_6}^2\},\notag\\
C_2\text{-PSG}\equiv& \text{group generated by } \{U_{T_1},U_{T_2},U_{C_2}\equiv U_{C_6}^3\},\notag\\
\text{transl.-PSG}\equiv& \text{group generated by } \{U_{T_1},U_{T_2}\},\label{eq:psg_subgroups}
\end{align}
where $C_6\text{-PSG}$ is the full-PSG described above. If the parton state respects the $C_6$-PSG, the two parton-boundary-condition phase factors $(\varphi_1,\varphi_2)$ are restricted to:
\begin{align}
\boldsymbol\varphi_{\rm FCI-a}=(\varphi_1,\varphi_2)_{\rm FCI-a}=\begin{cases}(0,0),&N=0\pmod 4,\\(\pi,\pi),&N=2\pmod 4.\end{cases}\label{eq:FCI_a_varphi}
\end{align}
where, for reasons that will become clear shortly, we denote this $C_6\text{-PSG}$ restricted PSG boundary condition with a subscript $_{\rm FCI-a}$.

On the other hand, if the parton state respects the $C_3$-PSG, these phase factors are more general and satisfy:
\begin{align}
\varphi_1=&-\varphi_2,& e^{i3\varphi_1}=&(-1)^{\frac{N(N-1)}{2}}.\label{eq:FCI_b_varphi}
\end{align}
Apart from the $\boldsymbol\varphi_{\rm FCI-a}$, up to a permutation of the two color, there is another solution, denoted with subscript $_{\rm FCI-b}$:
\begin{align}
\boldsymbol\varphi_{\rm FCI-b}=(\varphi_1,\varphi_2)_{\rm FCI-b}=\begin{cases}(\frac{2\pi}{3},\frac{4\pi}{3}),&N=0\pmod 4,\\(\frac{\pi}{3},\frac{5\pi}{3}),&N=2\pmod 4.\end{cases}\label{eq:FCI_b_varphi_1}
\end{align}
Eqs.~(\ref{eq:FCI_a_varphi}) and~(\ref{eq:FCI_b_varphi_1}) are essential to construct the twofold ground state sector in the FCI phase via Hdet.

\textbf{PSG exchanging color:} In the discussion above, we have assumed that the gauge transformation $W_g$ only involves phase factors $W_g=e^{i\phi_{g}(\mathbf r,\alpha)\tau^z}$. However, our numerical simulations show that, inside the FCI phase, for the variational Hdet saddle points, $W_g$ can involve color exchange: $i\tau_x\in SU(2)_g$. 

Let us focus on the FCI phase, which has a twofold topological ground-state degeneracy on a torus in the thermodynamic limit. For the standard modular action of a $C_6$ rotation on the ground-state manifold of the $\nu=1/2$ Laughlin topological order
\cite{Zhang2012}, the two many-body GS have $C_6$ eigenvalues that differ by a factor
of $e^{ i2\pi/3}$. Indeed, the exact diagonalization on the $N=4$ sample as well as the variational Hdet study on larger samples show that, one GS hosts $C_6=e^{\frac{i2\pi}{3}}$ (denoted as FCI-a) and the other GS hosts $C_6=1$ (denoted as FCI-b).

If one imposes the $C_6$-PSG on parton states, whose boundary condition is given by $\boldsymbol\varphi_{\rm FCI-a}$, we find that the optimal variational Hdet state always hosts $C_6=e^{\frac{i2\pi}{3}}$ inside the FCI phase, i.e., FCI-a, explaining the reason for the subscript in $\boldsymbol\varphi$. How to describe the other FCI-b state?

In fact, if one does not impose any rotation and only imposes the $\pi$-flux magnetic translation, i.e., transl.-PSG, the FCI-b state emerges in our variational Hdet simulation for the $N=4$ sample as a ground saddle point. For each individual color in the FCI-b variational ansatz, it satisfies only the $C_3$-PSG with boundary condition $\boldsymbol\varphi_{\rm FCI-b}$. To generate the full symmetry, one only needs to specify the projective implementation of $C_2$. After choosing a gauge consistent with Eq.~(\ref{eq:psg-phase}), we observe that the saddle point Hdet state of the FCI-b state satisfies:
\begin{align}
\text{FCI-b}: U_{C_2}=i\tau_x\cdot(-1)^x\cdot C_2,\label{eq:FCI-b-C_2-exchange}
\end{align}
exchanging the two colors. As a comparison, in the FCI-a state, $U_{C_2}$ is intra-color \footnote{Precisely, Eq.~(\ref{eq:FCI-a-C_2-internal}) is satisfied with the real-space PSG gauge-choice Eq.~(\ref{eq:psg-phase}) only for $N=0$ (mod 4) samples. On $N=2$ (mod 4) samples, Eqs.~(\ref{eq:FCI-b-C_2-exchange}) and~(\ref{eq:FCI-a-C_2-internal}) are satisfied after a real-space gauge transformation $G_\mathbf r\equiv e^{i\chi_N(\mathbf r)\tau_z}: \phi_g(\mathbf r,\alpha)\rightarrow \phi_g(\mathbf r,\alpha)+\chi_N(\mathbf r)-\chi_N(g^{-1}\mathbf r)$ from the gauge-choice Eq.~(\ref{eq:psg-phase}), where the $\chi_N$ phase on the u-sublattice is defined as $\chi_N(x,y,u)\equiv-\frac{\pi}{N}(x+y-\frac{3}{4})$ ($x,y=0,1...,N-1$), and its value on the v-sublattice is defined via $\chi_{N}(C_2(\mathbf r))\equiv-\chi_{N}(\mathbf r)$.}:
\begin{align}
\text{FCI-a}: U_{C_2}=U^3_{C_6}=i\tau_z\cdot(-1)^x \cdot C_2,\label{eq:FCI-a-C_2-internal}
\end{align}

Here, the factor $(-1)^x$ shared by the two $U_{C_2}$'s is due to the $\pi$-flux gauge choice in Eq.~(\ref{eq:psg-phase}). But the color exchange $i\tau_x$ in FCI-b is required simply because of the nontrivial boundary twist: Note that the parton mean-field state $|\Psi^{MF}(\boldsymbol\varphi)\rangle\equiv |\Psi^{(1)}_{\boldsymbol\varphi}\rangle\otimes |\Psi^{(2)}_{-\boldsymbol\varphi}\rangle$ (the sign flip of $\boldsymbol\varphi$ for two colors is from $\tau_z$). But under $C_2$, the boundary twist will change sign: $C_2:\boldsymbol\varphi\rightarrow -\boldsymbol\varphi$. Therefore, to transform the wavefunction back, a color exchange is required in $U_{C_2}$ for FCI-b. (Note that $\boldsymbol\varphi_{\rm FCI-a}=-\boldsymbol\varphi_{\rm FCI-a}$ but to $\boldsymbol\varphi_{\rm FCI-b}\neq -\boldsymbol\varphi_{\rm FCI-b}$).

\textbf{Emergent parton permutation symmetry inside the FCI phase:} We will use $|\Psi^{MF}_{\rm FCI-a}\rangle$ and $|\Psi^{MF}_{\rm FCI-b}\rangle$ to denote the two parton mean-field saddle points for the FCI-a and FCI-b states, respectively. Namely, $|\Psi^{MF}_{\rm FCI-a}\rangle$ satisfies the $C_6$-PSG with $\boldsymbol\varphi_{\rm FCI-a}$; $|\Psi^{MF}_{\rm FCI-b}\rangle$ satisfies the $C_3$-PSG with $\boldsymbol\varphi_{\rm FCI-b}$, together with Eq.~(\ref{eq:FCI-b-C_2-exchange}) implementing the $C_2$ rotation (denoted as FCI-b PSG).

Using the effective theory of Hdet states based on the variational manifold path integral \cite{Lin2026}, we know that the twofold GS manifold is captured by coupling the parton mean-field state $|\Psi^{MF}\rangle$ with flat connections of the $U(1)_g$ gauge field on the torus before the fusion, characterized by the twisted boundary conditions $\boldsymbol\varphi\equiv(\varphi_1,\varphi_2)$ in the PSG identity Eq.~(\ref{eq:psg_group_identities}). After fusion, due to the locality structure of the Hdet states, assuming the mean-field state $|\Psi^{MF}\rangle$ is short-range correlated, the expectation value of every local Hamiltonian term in  $|\Psi^{(b)}(\boldsymbol\varphi)\rangle\equiv \hat{\mathbf F}^{N_b}|\Psi^{MF}(\boldsymbol\varphi)\rangle$ is insensitive to $\boldsymbol\varphi$, up to exponentially small finite-size corrections. 

This is exactly the underlying relation between $|\Psi^{MF}_{\rm FCI-a}\rangle$ and $|\Psi^{MF}_{\rm FCI-b}\rangle$. In terms of the variational wavefunction, we expect that when $R$ is increased to a finite value, the Hdet wavefunction should be able to accurately capture the microscopic physics. In the thermodynamic limit, the saddle points $|\Psi^{MF}_{\rm FCI-a}\rangle$ and $|\Psi^{MF}_{\rm FCI-b}\rangle$ are not independent. \emph{They should differ only by a flat connection and are locally identical up to gauge redundancy.}

If the saddle point $|\Psi^{MF}_{\rm FCI-b}\rangle$ satisfies $U_{C_2}$ transformation in Eq.~(\ref{eq:FCI-b-C_2-exchange}), then after removing the flat connection $\vec {\mathsf A}$, the corresponding saddle point $|\Psi^{MF}_{\rm FCI-a}\rangle$ should also satisfy Eq.~(\ref{eq:FCI-b-C_2-exchange}) with $i\tau_x$. But we also know $|\Psi^{MF}_{\rm FCI-a}\rangle$ should also satisfy Eq.~(\ref{eq:FCI-a-C_2-internal}) with $i\tau_z$. There is only one possibility: $|\Psi^{MF}_{\rm FCI-a}\rangle$ need to satisfy the parton permutation symmetry:
\begin{align}
|\Psi^{(1)}_{\boldsymbol\varphi_{\rm FCI-a}}\rangle=|\Psi^{(2)}_{\boldsymbol\varphi_{\rm FCI-a}}\rangle,
\end{align}
\emph{at least in the thermodynamic limit, and when $R$ is large enough to capture the microscopic physics}. Namely, neither Eq.~(\ref{eq:FCI-b-C_2-exchange}) nor Eq.~(\ref{eq:FCI-a-C_2-internal}) alone enforces the parton color permutation symmetry, but together they do. We will show direct numerical evidence supporting this argument in Sec.~\ref{sec:simulation_results}.

As a sanity check, if one starts with a short-range parent mean-field Hamiltonian $h_{\rm FCI-a}$ satisfying the $C_6$-PSG \emph{and} color symmetry $(1)\leftrightarrow (2)$, after adding the flat connection:
\begin{align}
\vec {\mathsf A}\cdot\tau_z=\frac{1}{N}(\boldsymbol\varphi_{\rm FCI-b}-\boldsymbol\varphi_{\rm FCI-a})\cdot\tau_z \label{eq:flat_connection}
\end{align}
via Peierls substitution, indeed one will obtain a mean-field Hamiltonian $h_{\rm FCI-b}$ satisfying the FCI-b PSG up to gauge choice. 

The consequence of this argument is striking. The correct Hdet saddle points respecting $C_6$ symmetry should have emergent parton permutation symmetry. Algebraically, this enforces IGG$=SU(2)_g$ inside the whole FCI phase. Physically, this immediately rules out the traditional scenario of the FCI-SF continuous transition, where only one color closes its gap at the transition point leading to $N_f=2$ QED. Since IGG$=SU(2)_g$, both colors' gaps must close simultaneously. The most natural description for the phase transition is instead $N_f=2$ QCD.

\section{Classifying symmetry classes of Hdet states: representations of PSG and Chern-number constraints }\label{sec:classify_symmetry_classes}
For a given PSG, the mean-field state $|\Psi^{MF}\rangle=|\Psi^{(1)}\rangle\otimes|\Psi^{(2)}\rangle$ may still form different \emph{representations} of the PSG, which we coin as \textbf{symmetry classes}. For instance, the $C_6$-PSG is fully fixed by Eqs.~(\ref{eq:psg_group_identities}, \ref{eq:psg-phase}) in real space; however, the FCI-a and SF($\Gamma$) symmetry classes discussed below, both respect the same $C_6$-PSG, are inequivalent representations (see Table~\ref{tab:classes} for definitions).  Here, we show that different PSG symmetry classes constrain the Chern numbers $\mathcal C^{(p)}$ carried by the parton-$p$'s band structure in the thermodynamic limit. 

Assuming the parton mean-field states are short-range correlated, these Chern numbers qualitatively determine the quantum phase of the physical Hdet state in the thermodynamic limit based on the effective gauge theories (see section on the variational manifold path integral in Ref.~\cite{Lin2026} for details). As mentioned before, if $\mathcal C^{(1)}=\mathcal C^{(2)}=1$, the system feature a Laughlin-type topological order, i.e., a FCI phase. However if $\mathcal C^{(1)}=-\mathcal C^{(2)}=1$, the system is in a superfluid phase. Therefore, classifying the symmetry classes of the Hdet states allows one to construct symmetric variational ansatz describing specific quantum phases. \emph{Below we consider $C_6$-PSG as a demonstration.}

Because of Eq.~(\ref{eq:pi_flux_condition}), we choose each parton-$p$'s mean-field state $|\Psi^{(p)}\rangle$ as a Slater determinant formed by filling a single-particle band basis $\{|\phi^{(p)}_\mathbf k\rangle\}$ that are mutual eigenstates of $U_{T_1}$ and $(U_{T_2})^2$:
\begin{align}
U_{T_1} |\phi^{(p)}_\mathbf k\rangle=&e^{-i k_1} |\phi^{(p)}_\mathbf k\rangle,&(U_{T_2})^2 |\phi^{(p)}_\mathbf k\rangle=&e^{-2i k_2} |\phi^{(p)}_\mathbf k\rangle,
\end{align}
whose Brillouin Zone (BZ) can be chosen to be $(k_1,k_2)\in [0,2\pi)\times[0,\pi)$. It is convenient to define a reduced Brillouin zone (RBZ): $(k_1,k_2)\in [0,\pi)\times[0,\pi)$ and choose $|\phi^{(p)}_{\mathbf k+(\pi,0)}\rangle=U_{T_2} |\phi^{(p)}_{\mathbf k}\rangle$, $\forall \mathbf k\in {\rm RBZ}$. The doublet $\{ |\phi^{(p)}_{\mathbf k}\rangle,U_{T_2} |\phi^{(p)}_{\mathbf k}\rangle\}$ forms the twofold degenerate eigen-subspace of $(U_{T_1})^2,(U_{T_2})^2$ with eigenvalues $e^{-i 2k_1},e^{-i 2k_2}$. From Eq.~(\ref{eq:psg_group_identities}), one has:
\begin{align}
  U_{C_6}U_{T_1}^2U_{T_2}^{-2}=(-1)U_{T_1}^2U_{C_6},\;\;U_{C_6}U_{T_1}^2=U_{T_2}^2U_{C_6}.
\end{align}
Namely, $\forall \mathbf k=(k_1,k_2)\in {\rm RBZ}$, one could define a $C_6$ rotated $\mathbf k$-point: $ R\mathbf k=(Rk_1,Rk_2)$: 
\begin{align}
(Rk_1, Rk_2) &= (k_1-k_2+\frac{\pi}{2},k_1) \mod \pi,
\end{align}
so that the subspace spanned by $\{ |\phi^{(p)}_{\mathbf k}\rangle,U_{T_2} |\phi^{(p)}_{\mathbf k}\rangle\}$ is rotated into the subspace spanned by $\{ |\phi^{(p)}_{ R\mathbf k}\rangle,U_{T_2} |\phi^{(p)}_{R\mathbf k}\rangle\}$ under $U_{C_6}$. Therefore, the $C_6$ rotation center should be identified as $(\pi/2,\pi/2)$ (denoted as $\Gamma$ below) and the $U_{C_6}$ action on the filled band can be described by a 2-by-2 unitary matrix per $\mathbf k$-doublet:
\begin{align}
\mathsf R_{C_6}(\mathbf k)\equiv& \begin{pmatrix}\langle\phi^{(p)}_{R\mathbf k}|U_{C_6}|\phi^{(p)}_{\mathbf k}\rangle &\langle\phi^{(p)}_{R\mathbf k}|U_{C_6} U_{T_2}|\phi^{(p)}_{\mathbf k}\rangle\\ \langle\phi^{(p)}_{R\mathbf k}|U_{T_2}^{\dagger}U_{C_6}|\phi^{(p)}_{\mathbf k}\rangle &\langle\phi^{(p)}_{R\mathbf k}|U_{T_2}^{\dagger}U_{C_6} U_{T_2}|\phi^{(p)}_{\mathbf k}\rangle\end{pmatrix}\notag\\
=&\langle\phi^{(p)}_{R\mathbf k}|U_{C_6}|\phi^{(p)}_{\mathbf k}\rangle\begin{pmatrix}1&e^{i(Rk_1-k_1)}\\ e^{ik_1}&-e^{iRk_1}\end{pmatrix}.\label{eq:R_C6}
\end{align}
A similar analysis applies to $U_{C_3}$ and $U_{C_2}$.

\begin{figure}
    \centering
    \includegraphics[width=0.9\linewidth]{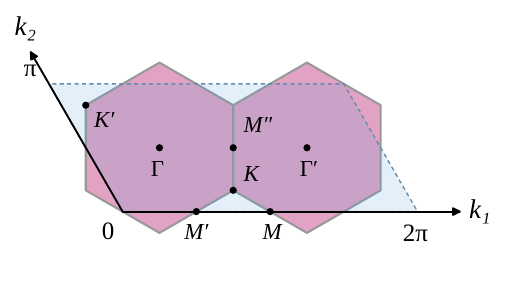}
    \caption{The parton's Brillouin Zone with high symmetry points marked.}
    \label{fig:parton_BZ}
\end{figure}

The parton-$p$'s \emph{representation} of the $C_6$-PSG boils down to the rotation matrix elements at the high-symmetry $\mathbf k$-points, whose symmetry-unrelated representatives can be chosen as (see Fig.~\ref{fig:parton_BZ}):
\begin{align}
\Gamma: (\frac{\pi}{2},\frac{\pi}{2}),\;\;K: (\frac{5\pi}{6},\frac{\pi}{6}), \;\; M: (\pi,0).
\end{align}
We define the band-basis-gauge-invariant indices as:
\begin{align}
\alpha^{(p)}_\Gamma\equiv&(-1)^{p-1}\cdot\sqrt 2\cdot\langle \phi^{(p)}_{\Gamma}|U_{C_6}|\phi^{(p)}_{\Gamma}\rangle\equiv e^{i\frac{4 a_{\Gamma}^{(p)}-1}{12}\pi},\notag\\
\beta^{(p)}_K\equiv&\sqrt 2\cdot \langle \phi^{(p)}_{K}|U_{T_2}U_{C_3}|\phi^{(p)}_{K}\rangle\equiv e^{i\frac{8 b_{K}^{(p)}+3}{12}\pi},\notag\\
\gamma^{(p)}_M\equiv&(-1)^{p-1}\cdot\langle \phi^{(p)}_{M}|U_{T_2}U_{C_2}|\phi^{(p)}_{M}\rangle\equiv  i(-1)^{c_M^{(p)}}.\label{eq:C6_psg_indices}
\end{align}
The factor $(-1)^{p-1}$ ($p=1,2$), i.e., $\tau_z$, in $\alpha$ and $\gamma$ is introduced to ensure that these indices are invariant under the color relabeling $(1)\leftrightarrow (2)$. In Appendix~\ref{app:psg_rep_derivation} we show that $a_{\Gamma}^{(p)}, b_{K}^{(p)},c_M^{(p)}$ are integers. \emph{Throughout this paper, we use $\alpha$, $\beta$, $\gamma$ to denote the $C_6$, $C_3$, $C_2$ gauge-invariant $U(1)$-phase indices, whose values are determined by integers $a,b,c$, modulo $6,3,2$, respectively.} Importantly, in the thermodynamic limit, they constrain the Chern number $\mathcal C^{(p)}$ (mod 12) of the parton state $|\Psi^{(p)}\rangle$ (see Appendix~\ref{app:psg_rep_derivation}):
\begin{align}
&C_6\text{-PSG: }e^{i2\pi \frac{\mathcal C^{(p)}}{2\cdot 6}}= e^{-i\frac{5\pi}{6}}\cdot\alpha^{(p)}_\Gamma  \beta_K^{(p)} \gamma^{(p)}_M ,\notag\\
&\text{i.e., } \mathcal C^{(p)}=2a_{\Gamma}^{(p)}+4b_K^{(p)}+6c_M^{(p)}-1 \pmod{12}. \label{eq:C6_mod_12_constraint}
\end{align}
This is a projective generalization \cite{FangCano2023} of the constraints relating rotation eigenvalues to the Chern number \cite{JianQi2013,Fang2012}. For the FCI-a state, since the $K$-point requires $N$ to be a multiple of $6$ due to Eq.~(\ref{eq:FCI_a_varphi}), one can obtain a weaker condition constraining $\mathcal C^{(p)}$ (mod 4) by taking the cubic power of Eq.~(\ref{eq:C6_mod_12_constraint}):
\begin{align}
&C_6\text{-PSG: }e^{i2\pi \frac{\mathcal C^{(p)}}{2\cdot 2}}=e^{i\frac{\pi}{4}}\cdot (\alpha^{(p)}_\Gamma)^3\cdot  (\gamma^{(p)}_M)^3,\notag\\
&\text{i.e., } \mathcal C^{(p)}=2a_{\Gamma}^{(p)}+2c_M^{(p)}-1 \pmod{4}. 
\label{eq:C6_mod_4_constraint}
\end{align}
which is applicable when $N$ is a multiple of 4.

If one considers the $C_3$-PSG or $C_2$-PSG, similar relations between rotation indices and the Chern number of the parton band can be established:
\begin{align}
&C_3\text{-PSG: }e^{i2\pi \frac{\mathcal C^{(p)}}{2\cdot 3}}=e^{-i\frac{\pi}{4}}\cdot\beta^{(p)}_\Gamma  \beta_K^{(p)}  \beta^{(p)}_{K'} ,\notag\\
&\text{i.e., } \mathcal C^{(p)}=2b_{\Gamma}^{(p)}+2b_K^{(p)}+2b_{K'}^{(p)}+1 \pmod{6}. \label{eq:C3_mod_6_constraint}
\end{align}
\begin{align}
&C_2\text{-PSG: }e^{i2\pi \frac{\mathcal C^{(p)}}{2\cdot 2}}=-i\cdot \gamma^{(p)}_\Gamma \gamma_M^{(p)} \gamma_{M'}^{(p)}\gamma_{M''}^{(p)} ,\notag\\
&\text{i.e., } \mathcal C^{(p)}=2c_{\Gamma}^{(p)}+2c_{M}^{(p)}+2c_{M'}^{(p)}+2c_{M''}^{(p)}+1 \pmod{4}. \label{eq:C2_mod_4_constraint}
\end{align}
We list the definition of relevant indices below (see Fig.~\ref{fig:parton_BZ}):
\begin{align}
K'\equiv&(\frac{\pi}{6},\frac{5\pi}{6}),\;\;M'\equiv (\frac{\pi}{2},0),\;\;M''\equiv (\pi,\frac{\pi}{2}),\notag\\
\beta^{(p)}_{\Gamma}\equiv&\sqrt 2\cdot\langle \phi^{(p)}_{\Gamma}|U_{C_3}|\phi^{(p)}_{\Gamma}\rangle\equiv e^{i\frac{8 b_{\Gamma}^{(p)}+1}{12}\pi},\notag\\
\beta^{(p)}_{K'}\equiv&\sqrt 2\cdot\langle \phi^{(p)}_{K'}|U_{T_2}U_{C_3}|\phi^{(p)}_{K'}\rangle\equiv e^{i\frac{8 b_{K'}^{(p)}+3}{12}\pi},\notag\\
\gamma_\Gamma^{(p)}\equiv&(-1)^{p-1}\cdot\langle \phi^{(p)}_{\Gamma}|U_{C_2}|\phi^{(p)}_{\Gamma}\rangle\equiv (-1)^{c_{\Gamma}^{(p)}},\notag\\
\gamma_{M'}^{(p)}\equiv&(-1)^{p-1}\cdot\langle \phi^{(p)}_{M'}|U_{C_2}|\phi^{(p)}_{M'}\rangle\equiv (-1)^{c_{M'}^{(p)}},\notag\\
\gamma_{M''}^{(p)}\equiv&(-1)^{p-1}\cdot\langle \phi^{(p)}_{M''}|U_{T_2}U_{C_2}|\phi^{(p)}_{M''}\rangle\equiv i(-1)^{c_{M''}^{(p)}},\label{eq:other_psg_indices}
\end{align}

\section{Constructing symmetry classes of Hdet states for FCI and superfluids}
\begin{table*}[t]
\caption{Symmetry classes of Hdet states subject to various PSGs together with their many-body $C_6$ quantum numbers. A rotation index in bold font indicates the momentum point is accessible in the $N=4$ sample. We also list the intended parton Chern numbers and their relevant constraint equations, where a bold font equation indicates that the constraint is well-defined on the $N=4$ sample.}
\label{tab:classes}
\scriptsize
\begin{tabular*}{\textwidth}{@{\extracolsep{\fill}}lllll@{}}
\toprule
Symmetry class & PSG  & PSG-Representation indices & Intended $(\mathcal C^{(1)},\mathcal C^{(2)})$ & Physical $C_6$ sector\\
\midrule
FCI-a & $C_6$-PSG with $\boldsymbol\varphi_{\rm FCI-a}$  & $(\boldsymbol a_{\Gamma}^{(1)},b^{(1)}_{K},\boldsymbol c^{(1)}_M)=(\boldsymbol a_{\Gamma}^{(2)},b^{(2)}_{K},\boldsymbol c^{(2)}_M)=(0,2,1)$; & $(+1,+1)$, Eq.~(\ref{eq:C6_mod_12_constraint}) [\textbf{Eq.~(\ref{eq:C6_mod_4_constraint})}] & $e^{i2\pi/3}$\\
\midrule
FCI-b & $C_3$-PSG with $\boldsymbol\varphi_{\rm FCI-b}$  & $(b_{\Gamma}^{(1)},\boldsymbol b^{(1)}_{K}, b^{(1)}_{K'})=(b_{\Gamma}^{(2)}, b^{(2)}_{K},\boldsymbol b^{(2)}_{K'})=(0,2,1)$;& $(+1,+1)$, Eq.~(\ref{eq:C3_mod_6_constraint}) & $1$\\
&+exchange-$C_2$\\
\midrule
SF$(\Gamma)$ & $C_6$-PSG with $\boldsymbol\varphi_{\rm FCI-a}$  & $(\boldsymbol a_{\Gamma}^{(1)},b^{(1)}_{K},\boldsymbol c^{(1)}_M)=(0,2,1)$; $(\boldsymbol a_{\Gamma}^{(2)},b^{(2)}_{K},\boldsymbol c^{(2)}_M)=(-1,2,1)$; & $(+1,-1)$, Eq.~(\ref{eq:C6_mod_12_constraint}) [\textbf{Eq.~(\ref{eq:C6_mod_4_constraint})}] & $1$\\
\midrule
SF$(M)$ & $C_2$-PSG with $\boldsymbol\varphi_{\rm FCI-a}$  & $(\boldsymbol c_{\Gamma}^{(1)},\boldsymbol c^{(1)}_{M},\boldsymbol c^{(1)}_{M'},\boldsymbol c^{(1)}_{M''})=(0,0,0,0)$; & $(+1,-1)$, \textbf{Eq.~(\ref{eq:C2_mod_4_constraint})} & $C_6$ broken\\
& &$(\boldsymbol c_{\Gamma}^{(2)},\boldsymbol c^{(2)}_{M},\boldsymbol c^{(2)}_{M'},\boldsymbol c^{(2)}_{M''})=(0,1,0,0)$;& &\\
\bottomrule
\end{tabular*}
\end{table*}
In the next section, we will discuss simulation results for Hdet states in various symmetry classes, mainly focusing on the $N=4$ sample with the number of fusion channels $R=1$ and $R=3$. We also present more details for simulation results for larger samples $N=6,8,10,12$ with $R=1$ in Appendix~\ref{app:larger_VMC}. To prepare those discussions, we present here four specific symmetry classes of Hdet variational states relevant for the investigated phase diagram: FCI-a, FCI-b, SF($\Gamma$), and SF($M$), which are listed in Table~\ref{tab:classes}.

As explained before, FCI-a and FCI-b symmetry classes are aimed at the twofold degenerate ground states in the FCI phase whose many-body $C_6$ eigenvalue are $e^{i\frac{2\pi}{3}}$ and $1$, respectively. Indeed, these two symmetry classes are found to be the variational optimal Hdet states inside the FCI phase, carrying Chern numbers $\mathcal C^{(1)}=\mathcal C^{(2)}=1$. Importantly, their PSG representation indices are consistent with being related by a flat connection $\vec{\mathsf A}\cdot\tau_z$ as in Eq.~(\ref{eq:flat_connection}). Precisely, it is easy to show that restricting $C_6$-PSG to $C_3$-PSG induces a map of indices: 
\begin{align}
&\beta^{(p)}_{\Gamma}=e^{i\frac{\pi}{4}}(\alpha_\Gamma^{(p)})^2,\;\; \beta^{(p)}_{K'}=e^{-i\frac{2\pi}{3}}\beta^{(p)}_{K},\notag\\
&\text{i.e., } (b_{\Gamma}^{(p)},b_{K}^{(p)},b_{K'}^{(p)})=(a_{\Gamma}^{(p)}\text{ mod } 3, b_{K}^{(p)}, b_{K}^{(p)}-1 \text{ mod } 3),
\end{align}
and adding a flat connection will not change it.

SF($\Gamma$) and SF($M$) symmetry classes are aimed at the two superfluid phases, whose parton mean-field states carry $\mathcal C^{(1)}=-\mathcal C^{(2)}=1$. Here, the SF($\Gamma$) symmetry class respects $C_6$-PSG since the corresponding superfluid phase is symmetric. On the other hand, since the SF($M$) phase breaks the $C_6$ rotation symmetry, we require the corresponding symmetry class to respect only $C_2$-PSG, with $\boldsymbol\varphi_{\rm FCI-a}$ boundary twist. 

In particular, the putative continuous FCI-SF($\Gamma$) phase transition, in terms of the corresponding symmetry classes, is consistent with a Dirac band inversion for parton color-$(2)$ at a pair of magnetic-translation-related momentum points: $\Gamma=(\frac{\pi}{2},\frac{\pi}{2})$ and $\Gamma'\equiv (\frac{3\pi}{2},\frac{\pi}{2})$, where the symmetry index $a^{(2)}$ changes. Together, the Chern number change of $\mathcal C^{(2)}$ is $-2$. However, as emphasized before, this transition should features gap-closing of both parton colors simultaneously: the color-$(2)$ has a band inversion, while the color-$(1)$ has a trivial band-gap reopening.

In practice, we construct the variational Hdet states for each class as follows. For a given PSG (e.g., $C_6$-PSG or $C_3$-PSG), the symmetry transformation $U_{g}$ for the real-space parton-$p$'s orbitals $|(\mathbf r,\alpha)^{(p)}\rangle$ is given in Eq.~(\ref{eq:psg-phase}). The filled parton band basis $\{|\phi^{(p)}_{\mathbf k}\rangle\}$ is characterized by the overlap matrix 
\begin{align}
A^{(p)}_{(\mathbf r,\alpha);\mathbf k}\equiv \langle (\mathbf r,\alpha)^{(p)}|\phi^{(p)}_{\mathbf k}\rangle,\label{eq:overlap_mat_def}
\end{align}
with column index $\mathbf k$ and row index $(\mathbf r,\alpha)$, satisfying:
\begin{align}
\mathsf U^{(p)}_g \cdot A^{(p)}=& A^{(p)}\cdot \mathsf R^{(p)}_g, \;\;\forall U_g\in \text{PSG},
\label{eq:A_p_psg_condition}
\end{align}
where $\mathsf U^{(p)}_g$ is an $(N_{\rm site}\cdot R)$-dimensional square matrix defined by the real-space transformation $U_{g}$:
\begin{align}
[\mathsf U^{(p)}_g]_{(\mathbf r,\alpha),(\mathbf r',\alpha')}\equiv&\langle (\mathbf r,\alpha)^{(p)}| U_g|(\mathbf r',\alpha')^{(p)}\rangle.
\end{align}

On the other hand $\mathsf R^{(p)}_g$ is a $N_{b}$-dimensional square matrix defined by how the filled band basis $\{|\phi^{(p)}_{\mathbf k}\rangle\}$ transform, e.g. Eq.~(\ref{eq:R_C6}), which contains the PSG-representation indices introduced in Eqs.~(\ref{eq:C6_psg_indices}) and~(\ref{eq:other_psg_indices}). Note that general elements of $\mathsf R^{(p)}_g$ depend on the gauge choice of the band basis, i.e., $A^{(p)}$:
\begin{align}
\mathsf R_g=& A^{(p)+}(\mathsf U^{(p)}_g)A^{(p)}, \;\;\forall U_g\in \text{PSG},\label{eq:fixing_R_g}
\end{align}
where $^{+}$ is the Moore-Penrose pseudoinverse. If the band basis is orthonormal, $A^{(p)+}=A^{(p)\dagger}$. One can always choose a representative $A^{(p)}_{\rm rep}$ that realizes a specific set of band-basis-gauge-invariant representation indices, and use Eq.~(\ref{eq:fixing_R_g}) to fully fix $\mathsf R_g$, corresponding to fixing a gauge of the band basis. After $\mathsf R_g$ is fixed, the PSG condition Eq.~(\ref{eq:A_p_psg_condition}) constrains $A^{(p)}$ as a set of linear equations, leading to the PSG-symmetric variational subspace of the band basis, i.e., $A^{(p)}$ matrices.

Special consideration is needed for the FCI-b symmetry class, where $U_{C_2}$ rotation is implemented as Eq.~(\ref{eq:FCI-b-C_2-exchange}). There one first uses $C_3$-PSG with $\boldsymbol\varphi_{\rm FCI-b}$ to constrain the parton-$(1)$ state $A_{\boldsymbol\varphi_{\rm FCI-b}}^{(1)}$ as in Eq.~(\ref{eq:A_p_psg_condition}). Next, one can use $U_{C_2}$ in Eq.~(\ref{eq:FCI-b-C_2-exchange}) to generate the parton-$(2)$ state $A_{-\boldsymbol\varphi_{\rm FCI-b}}^{(2)}\equiv \mathsf U_{C_2}A_{\boldsymbol\varphi_{\rm FCI-b}}^{(1)}$. Note that, due to the basis-independence theorem in Ref.~\cite{Lin2026}, the column rearrangement (i.e., the ordering of band-basis $\mathbf k$) of the $A^{(p)}$ matrix does not change the Hdet state after fusion.

For each set of overlap matrices $\{A^{(1)},A^{(2)}\}$, due to  the fusion operator Eq.~(\ref{eq:Hdet_canonical_parton_construction}), the full fusion tensor $T$ in Eq.~(\ref{eq:CPD}) is obtained:
\begin{align}
T_{\mathbf r,\mathbf k,\mathbf k'}\equiv \sum_{\alpha=1}^R A^{(1)}_{(\mathbf r,\alpha),\mathbf k}A^{(2)}_{(\mathbf r,\alpha),\mathbf k'},
\end{align}
and the physical boson state follows Eq.~(\ref{eq:Hdet_def})

A subtlety arises when $R>1$: The overlap-matrix $\{A^{(1)}, A^{(2)}\}$ has a local $GL(R,\mathbb{C})$ gauge redundancy. Let $A^{(p)}_{\mathbf r}$ denote the $R\times N_b$ block of $A^{(p)}$ at site $\mathbf r$. Within the color-separated parametrization, the transformations
\begin{align}
A^{(1)}_{\mathbf r}\longrightarrow&
G_{\mathbf r}A^{(1)}_{\mathbf r},\notag\\
A^{(2)}_{\mathbf r}\longrightarrow&
G_{\mathbf r}^{-T}A^{(2)}_{\mathbf r},
\;\;G_{\mathbf r}\in GL(R,\mathbb{C}),
\label{eq:local_GL_redundancy}
\end{align}
where $GL(R,\mathbb{C})$ is the group of invertible complex $R\times R$ matrices, leave the fusion tensor and the physical state invariant:
\begin{align}
T[\mathbf r]\equiv&
(A^{(1)}_{\mathbf r})^T A^{(2)}_{\mathbf r},\notag\\
T[\mathbf r]\longrightarrow&
(A^{(1)}_{\mathbf r})^T
G_{\mathbf r}^T G_{\mathbf r}^{-T}
A^{(2)}_{\mathbf r}
=T[\mathbf r],
\label{eq:tensor_GL_inv}
\end{align}
where $T[\mathbf r]$ is the matrix obtained by fixing $\mathbf r$ in the fusion tensor $T_{\mathbf r,\mathbf k,\mathbf k'}$.

We fix the noncompact $GL(R,\mathbb{C})$ gauge redundancy as follows.
We begin by defining the onsite one-body density matrices:
\begin{align}
[\rho^{(p)}_{\mathbf r}]_{\alpha\beta}\equiv&
\left\langle
f^{(p)\dagger}_{\mathbf r,\beta}
f^{(p)}_{\mathbf r,\alpha}
\right\rangle_{\rm MF}\notag\\
=&
[A^{(p)}A^{(p)+}]_{(\mathbf r,\alpha),(\mathbf r,\beta)}.
\label{eq:parton_onsite_density_matrix}
\end{align}
Here $A^{(p)}A^{(p)+}$ denotes the occupied-state projector,
which reduces to $A^{(p)}A^{(p)\dagger}$ when the occupied
states are orthonormal.

One natural way to partially fix $GL(R,\mathbb{C})$ redundancy to avoid its non-compact directions is to find a parton
mean-field state with minimal norm along the local $GL(R,\mathbb{C})$ gauge orbit; see Appendix~\ref{app:min_mf}.
The resulting condition is
\begin{align}
\rho^{(1)}_{\mathbf r}-(\rho^{(2)}_{\mathbf r})^T=0,
\qquad \forall \mathbf r.
\label{eq:zero-moment-map condition}
\end{align}
Starting from a generic representative
$\{A^{(1)},A^{(2)}\}$, we employ a Newton-based line-search
algorithm to find local $GL(R,\mathbb{C})$ gauge transformations
$G_{\mathbf r}$ that yield a mean-field state satisfying
Eq.~(\ref{eq:zero-moment-map condition}). This condition is equivalent to the local $U(R)$ gauge-charge
neutrality condition imposed at the mean-field level, where $U(R)$ is the maximal compact subgroup of $GL(R,\mathbb{C})$:
\begin{align}
\langle Q^a_{\mathbf r}\rangle
\equiv
{\rm Tr}\!\left[
t^a\big(\rho^{(1)}_{\mathbf r}
-(\rho^{(2)}_{\mathbf r})^T\big)
\right]=0.
\label{eq:local_U_R_gauge_charges}
\end{align}
Here, $t^a$ are generators of $U(R)$. Eq.~(\ref{eq:local_U_R_gauge_charges}) shares the same form as the gauge neutrality constraints imposed by Lagrange multipliers in the field theory treatments of the saddle points in parton constructions.

Once Eq.~(\ref{eq:local_U_R_gauge_charges}) is imposed,
the residual local gauge redundancy is $U(R)$, which is partially fixed by the PSG Eq.~(\ref{eq:A_p_psg_condition}).
One can show that the parton's Berry flux
through each momentum-space plaquette (see Eq.~(\ref{eq:berry_flux})) is invariant under this remaining $U(R)$ gauge redundancy.

\section{Simulation Results}\label{sec:simulation_results}
\textbf{Outline:} Throughout this paper, we consider rotation-symmetric samples of the honeycomb lattice with $N\times N$ unit cells with $N_b=\frac{N^2}{2}$ hard-core bosons. We have performed the variational Hdet simulation for the phase diagram of the model in Eq.~(\ref{eq:hard_core_boson_model}) by tuning $t''$ only while fixing $t=1,t'=0.6,\phi=0.4\pi$ and $V_1=V_2=0$. For each symmetry class, the optimal Hdet state is obtained by optimizing the symmetry-allowed overlap matrices $A^{(p)}$ satisfying Eq.~(\ref{eq:A_p_psg_condition}). 

\begin{figure}
    \centering
    \includegraphics[width=1.0\linewidth]{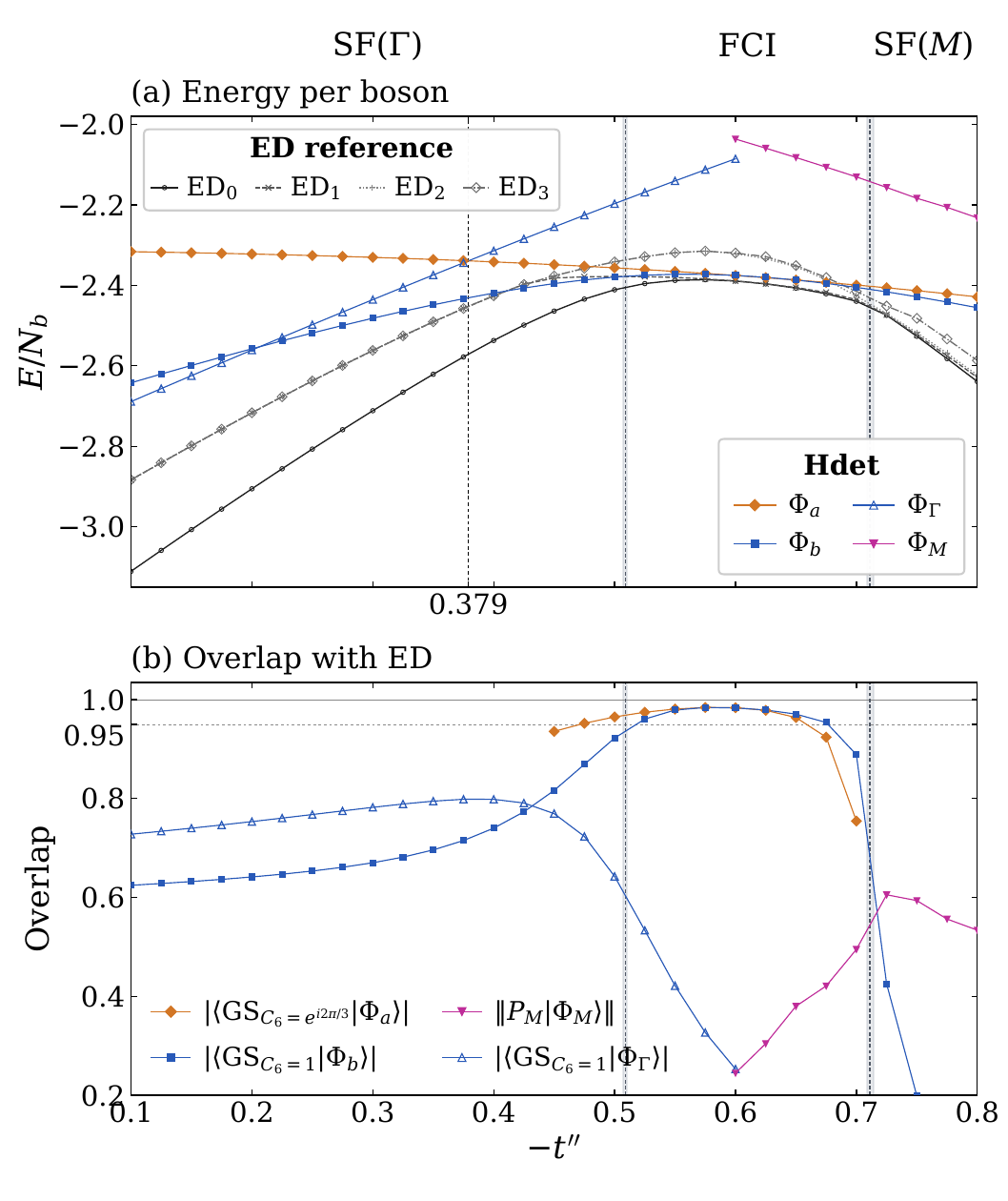}
    \caption{For the $N=4$ sample, we present the optimal $R=1$ Hdet states' variational energy per boson for different symmetry classes in (a), where the lowest four ED eigenlevels ($\text{ED}_0$ is the lowest) are also shown as references. In (b), their overlap with ED eigenstates are shown: $|GS_{C_6=e^{i\frac{2\pi}{3}}}\rangle$ and $|GS_{C_6=1}\rangle$ are defined as the ED eigenstates in the specific $C_6$ sectors among the two lowest levels. $P_M\equiv \sum_{i=0}^2|\text{ED}_i\rangle \langle \text{ED}_i|$ is defined to be the projector into the lowest three ED eigenstates. The Hdet state $|\Phi_M\rangle$ breaks $C_6$ and has nonzero overlap with all three nearly degenerate ground states in the SF($M$) phase.}
    \label{fig:N4R1}
\end{figure}

\begin{figure}
    \centering
    \includegraphics[width=1.0\linewidth]{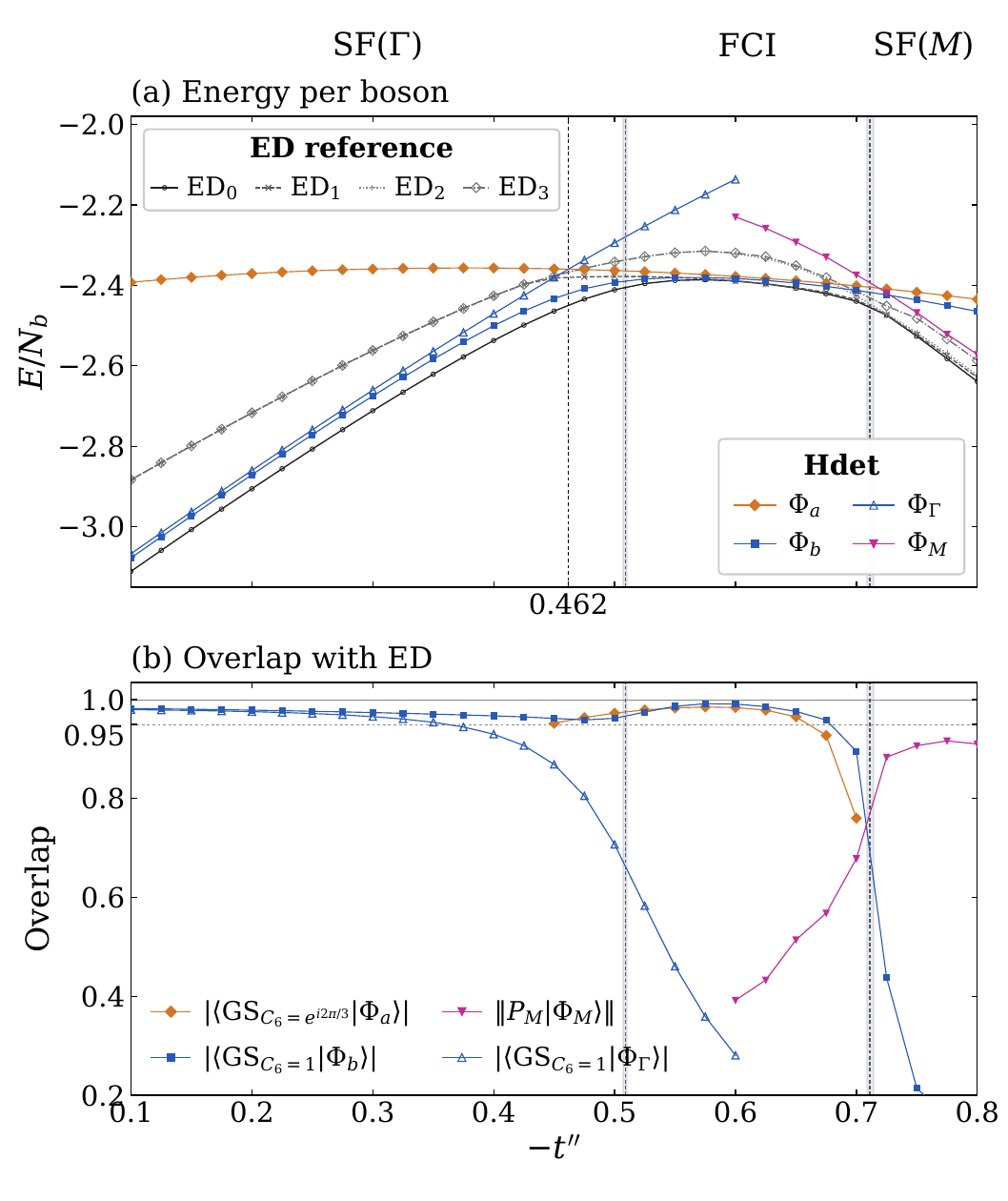}
    \caption {Similar plots to Fig.~\ref{fig:N4R1}, but for the optimal $R=3$ Hdet states on the $N=4$ sample. The improvement of energetics and overlaps are significant, particularly in the superfluids.}
    \label{fig:N4R3}
\end{figure}

To investigate the difference from the traditional parton construction Eq.~(\ref{eq:traditional_parton_construction}) with $R=1$, we simulate the $N=4$ sample with 8 bosons with $R=1$ and $R=3$. The optimal energies per boson as well as the overlaps with the ED ground states are plotted in Fig.~\ref{fig:N4R1} and Fig.~\ref{fig:N4R3} for $R=1$ and $R=3$, respectively. Here, for $R=1$, due to the small system size, both the optimization and the reported measurements (energy, overlap, etc.) are performed exactly. For $R=3$, although the reported measurements are exact, the optimization is performed using Monte Carlo estimates
of the gradients, with a combination of the Adam optimizer and
stochastic reconfiguration \cite{Sorella2001,Sorella2005,BeccaSorella2017}. As a benchmark reference, we also performed the exact diagonalization (ED) for the $N=4$ sample and computed the overlaps between the Hdet optimal states and the ED ground states (GS).

In addition to $N=4$, we also performed variational Monte Carlo (VMC) simulations for $R=1$ Hdet states on larger samples with $N=6,8,10,12$, where optimization is again performed using the stochastic gradient descent method. The detailed simulation results for these larger samples are presented in Appendix~\ref{app:larger_VMC}. Unfortunately, $R>1$ VMC simulation is not feasible for these larger system sizes since exactly computing $R>1$ Hdet is NP-hard. One way to make progress on large system sizes with $R>1$ is to apply the projective-expansion method introduced in Ref.~\cite{Lin2026}, which can simulate Hdet states with $R>1$ under an approximation scheme improvable order by order. We leave the projective expansion study as a future project.

We will also present the Berry curvature distribution for the parton's mean-field states. For parton color-$(p)$, we define the Berry flux $\boldsymbol\phi^{(p)}_\mathbf p$ in a discrete momentum \emph{triangular} plaquette $\mathbf p$ as follows. Let $\mathbf k_1,\ldots,\mathbf k_3$ denote the plaquette corners ordered counterclockwise, with $\mathbf k_4=\mathbf k_1$,
\begin{align}
\boldsymbol\phi^{(p)}_\mathbf p
&\equiv \text{Arg}\Big[
\prod_{j=1}^{3}
\left\langle \phi^{(p)}_{\mathbf k_{j+1}} \right|
e^{i\Delta\mathbf k_j\cdot\hat{\mathbf r}}
\left| \phi^{(p)}_{\mathbf k_j} \right\rangle\Big],\label{eq:berry_flux}
\end{align}
where $\Delta\mathbf k_j\equiv \mathbf k_{j+1}-\mathbf k_j$, and $\hat{\mathbf r}$ is the position operator.

\textbf{ED results: } First of all, by increasing $-t''$ from $0.1$ to $0.8$, we observe that the ED spectra are consistent with a continuous phase transition between the SF($\Gamma$) phase and the FCI phase near $-t''\sim 0.5$ followed by a first-order phase transition from the FCI phase and the SF($M$) phase at $-t''\sim 0.7$. The SF($\Gamma$) phase features a unique GS with $C_6=1$ eigenvalue; the FCI phase features a nearly twofold degenerate GS with $C_6=1,e^{i\frac{2\pi}{3}}$ eigenvalues; and the SF($M$) phase features a nearly three-fold degenerate GS with $C_6=1,e^{\pm i\frac{2\pi}{3}}$ eigenvalues. All these GS are found to be in the zero center-of-momentum sector $COM=(0,0)$. 

\begin{figure}
    \centering
    \includegraphics[width=0.9\linewidth]{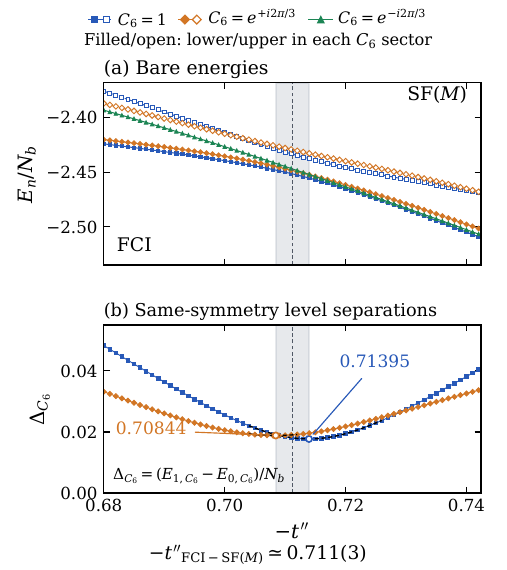}
    \caption{(a) $ C_6$-sector-resolved low-energy ED spectra of the $N=4$ sample near the FCI-SF($M$) phase boundary, which is consistent with avoided level crossings for the $C_6=e^{i\frac{2\pi}{3}}$ and $C_6=1$ sectors. (b) The same $C_6$-sector level separations show that the minima of the separation occur at slightly different $t''$'s for the two sectors, and their average is the estimated $t''_{\text{FCI-SF($M$)}}$.}
    \label{fig:ED_FCI_SFM_boundary}
\end{figure}

Although two of the three GS in the SF($M$) phase share symmetry quantum numbers with the two GS in the FCI phase, the FCI-SF($M$) transition can be well understood as an avoided level crossing at $-t''_{\text{FCI-SF($M$)}}= 0.711(3)$ for these two sectors, as shown in Fig.~\ref{fig:ED_FCI_SFM_boundary}, which is consistent with a first-order phase transition. On the other hand, the $C_6=1$ GS evolves smoothly over the SF($\Gamma$)-FCI phase transition. 

\begin{figure}
    \centering
    \includegraphics[width=0.9\linewidth]{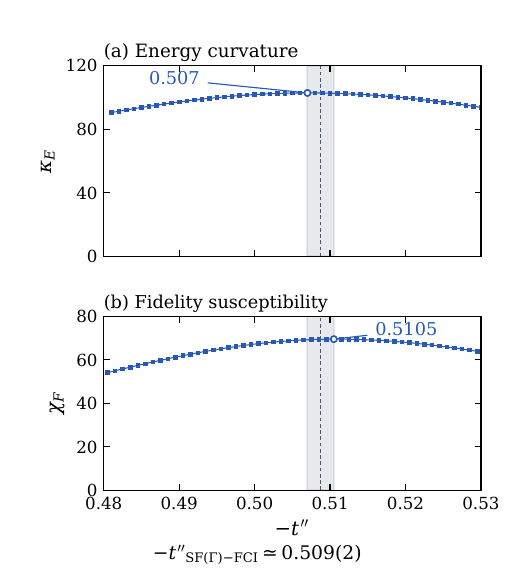}
    \caption{Estimating the SF($\Gamma$)--FCI boundary by identifying the peaks in (a): the energy curvature and (b): the fidelity susceptibility based on $N=4$ ED results. (a): Curvature of the total ground-state energy is defined as $\kappa_E\equiv-d^2E_{\mathrm{GS}}/d(t'')^2$. (b): Fidelity susceptibility \cite{You2007} is defined as $\chi_{F,j}\equiv-2\ln F_j/\delta^2$, where the overlap of adjacent ground states $F_j\equiv |\langle\psi_{j-1}|\psi_j\rangle|$ with a $t''$-spacing $\delta\equiv |t''_j-t''_{j-1}|=0.001$. The positions of the peaks in (a) and (b) slightly differ, and their average is the estimated $t''_{\text{SF($\Gamma$)-FCI}}$.}
\label{fig:ED_SFGamma_FCI_boundary}
\end{figure}

Different from the first-order FCI-SF($M$) transition boundary, the boundary for the SF($\Gamma$)-FCI transition cannot be sharply determined due to the putative continuous nature and the current small sample size. Nevertheless, we performed the GS energy-curvature (i.e., the second derivative of the GS energy $\frac{d^2E_{GS}}{d(t'')^2}$) as well as the GS fidelity susceptibility \cite{You2007} on this sample, and estimate that $-t''_{\text{SF($\Gamma$)-FCI}}= 0.509(2)$ (see Fig.~\ref{fig:ED_SFGamma_FCI_boundary}). Both $t''_{\text{SF($\Gamma$)-FCI}}$ and $t''_{\text{FCI-SF($M$)}}$ are consistent with previously reported values based on DMRG studies on larger system sizes.

Next, we will move on to the variational results for the Hdet states. To distinguish them from the ED states, we label the variational Hdet states as $|\Phi_{a}\rangle, |\Phi_b\rangle, |\Phi_\Gamma\rangle,|\Phi_M\rangle$ for the FCI-a,  FCI-b, SF($\Gamma$), and SF($M$) symmetry classes, respectively. 

\textbf{$N=4$, $R=1$ Hdet -- energy and overlap:} In Fig.~\ref{fig:N4R1}, we observe that the FCI symmetry classes $|\Phi_{a}\rangle,|\Phi_{b}\rangle$ Hdet states perform well inside the FCI phase, giving accurate energetics as well as high overlaps with ED eigenstates. The drop of their overlaps near the FCI-SF($M$) transition is consistent with the avoided level-crossings observed in ED (see Fig.~\ref{fig:ED_FCI_SFM_boundary}).

However, for the SF symmetry classes, either $|\Phi_\Gamma\rangle$ or $|\Phi_M\rangle$ performs quite poorly for $R=1$, with a maximal overlap with ED $<0.8$ and significantly higher energies than ED. In particular, in the investigated regime, $|\Phi_M\rangle$ is always higher in energy than the FCI symmetry classes, and consequently the $R=1$ VMC results fail to capture the first-order FCI-SF($M$) transition. 

\textbf{$N=4$, $R=3$ Hdet -- energy and overlap:} As shown in Fig.~\ref{fig:N4R3}, we find that the improvement in both energy and overlap from $R=1$ to $R=3$ is modest for the FCI symmetry classes. On the contrary, the $R=3$ SF symmetry states improve drastically compared with $R=1$ Hdet states. For example, $|\Phi_\Gamma\rangle$ shows high overlap $>0.95$ with ED deep inside the SF($\Gamma$) phase. In addition, $|\Phi_M\rangle$ becomes the lowest-energy variational ansatz for $-t''>0.73$, successfully capturing the first-order FCI-SF($M$) transition. These findings show a major improvement in microscopics for the $R>1$ Hdet states compared with the traditional $R=1$ construction.

As a finite-size $N=4$ sample of a continuous phase transition, the $C_6=1$ ED GS wavefunction should be smoothly connected near the SF($\Gamma$)-FCI phase transition. Therefore, we do expect that the $|\Phi_{b}\rangle$ ansatz can capture the continuous change of the $C_6=1$ GS wavefunction near this transition. Indeed, $|\Phi_{b}\rangle$ variational energy shows a smooth curve near the transition, quantitatively close to the ED energy curve. But surprisingly, we find that the $|\Phi_{b}\rangle$ ansatz performs extremely well even deep inside the SF($\Gamma$) phase, even outperforming $|\Phi_\Gamma\rangle$. How to understand this observation?

First of all, this could be a finite-size effect. As shown in Appendix~\ref{app:larger_VMC}, at larger $N$ but $R=1$, $|\Phi_\Gamma\rangle$ clearly outperforms $|\Phi_{b}\rangle$ in the SF($\Gamma$) regime. It is still possible, however, that $|\Phi_{b}\rangle$ performs excellently in the SF($\Gamma$) phase when $R>1$. 

Note that in the thermodynamic limit, assuming the parton mean-field states are short-range correlated (i.e. gapped), the effective theories of $|\Phi_{b}\rangle$ and $|\Phi_\Gamma\rangle$ are fundamentally different: $|\Phi_{b}\rangle$ should carry parton Chern number $\mathcal C^{(1)}=\mathcal C^{(2)}=1$ giving Laughlin-type topological order while $|\Phi_\Gamma\rangle$ should carry $\mathcal C^{(1)}=-\mathcal C^{(2)}=1$ giving a Maxwell gauge dynamics (i.e. Goldstone mode of the superfluid). 

\begin{figure}
    \centering
    \includegraphics[width=0.8\linewidth]{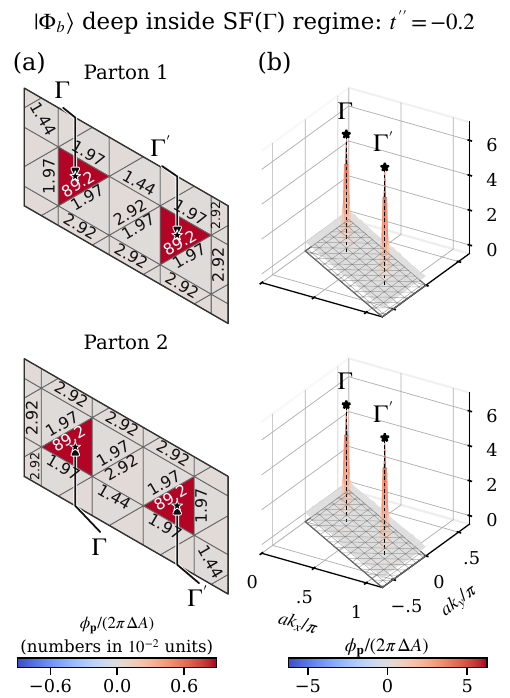}
    \caption{Parton's Berry curvature densities $\frac{1}{2\pi}\frac{\boldsymbol\phi^{(p)}_{\mathbf p}}{\Delta A_\mathbf p}$ of (a) $N=4$, $R=3$ (heatmap) and (b) $N=12$, $R=1$ (surface plot) optimal FCI-b Hdet states $|\Phi_b\rangle$ deep inside the SF($\Gamma$) phase ($t''=-0.2$).  $\boldsymbol\phi^{(p)}_{\mathbf p}$ is the Berry flux for the momentum space plaquette $\mathbf p$ for the parton color-$(p)$; $\Delta A_{\mathbf p}$ is the area of that plaquette. The Berry curvature is highly concentrated near the would-be band inversion points: $\Gamma$ and $\Gamma'$.}
    \label{fig:FCIb_in_SF}
\end{figure}

\begin{figure}
    \centering
    \includegraphics[width=0.8\linewidth]{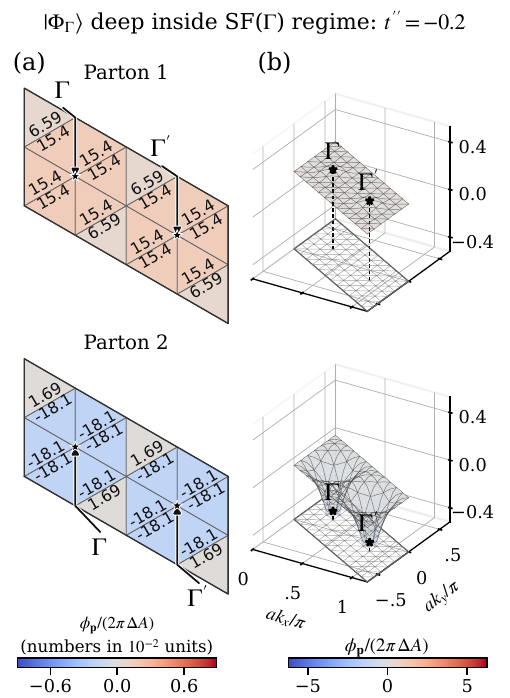}
    \caption{Similar to plots in Fig.~\ref{fig:FCIb_in_SF}, but for the optimal $|\Phi_\Gamma\rangle$ states, whose partons' Berry curvature is smoothly distributed.}
    \label{fig:FCI_Gamma_in_SF}
\end{figure}

As a constrained variational ansatz, what would $|\Phi_{b}\rangle$ behave deep inside the SF($\Gamma$) phase? Based on the effective theories, we expect that, when the system size is large enough, the parton mean-field states for $|\Phi_{b}\rangle$ are actually gapless inside the SF($\Gamma$) phase. Due to the symmetry class constraint (i.e., $C_2$-exchange PSG in Eq.~(\ref{eq:FCI-b-C_2-exchange})), a true one-color band inversion is not possible since the two parton colors' mean-field states are related by $C_2$, and a gapless $|\Phi_{b}\rangle$ is the best approximation towards the SF($\Gamma$) phase. Indeed, we plot the Berry curvature distribution of the optimal $|\Phi_{b}\rangle$'s parton mean-field states in the SF($\Gamma$) regime at $t''=-0.2$ in Fig.~\ref{fig:FCIb_in_SF}, for $N=4$, $R=3$ and $N=12$, $R=1$, respectively. Clear peaks of the Berry curvature are observed at the would-be band-inversion points: $\Gamma$ and $\Gamma'$, consistent with this expectation. For comparison, in Fig.~\ref{fig:FCI_Gamma_in_SF}, we present the Berry curvature distribution for the optimal $|\Phi_{\Gamma}\rangle$'s parton mean-field states at $t''=-0.2$, which is smooth. 

Therefore, based on the variational manifold path integral \cite{Lin2026}, the effective theory for the quantum fluctuations near the $|\Phi_{b}\rangle$ saddle point inside the SF($\Gamma$) phase would be gapless fermions coupling with a dynamical gauge field. This is most likely inconsistent with the correct long-range physics: a free boson theory (the Goldstone mode), which the $|\Phi_{\Gamma}\rangle$ saddle point captures well via the Maxwell photon.

Even if the excellent performance of the $|\Phi_{b}\rangle$ ansatz inside the SF($\Gamma$) phase is not a finite-size effect, our current interpretation is analogous to a situation known in the tensor-network community. There, a projected entangled pair state (PEPS) with a finite bond dimension $D$ cannot truly describe a gapped Chern insulator due to a known obstruction \cite{Wahl2013,DubailRead2015}. If one insists on using PEPS as a variational ansatz to approximate a gapped Chern insulator, a gapless PEPS can be the optimal state. That optimal PEPS can have extremely accurate energetics, a local physical property, but will fail for long-range physics. $|\Phi_{b}\rangle$ ansatz inside the SF($\Gamma$) phase is in a similar situation: it may have very good energetics, but is unable to capture the correct long-range physics.

\textbf{Evidence for emergent color-permutation symmetry in the FCI phase: } In Sec.~\ref{sec:psg}, we have argued that, based on the PSG analysis, the parton color-permutation symmetry is expected to emerge in the FCI phase for $|\Phi_a\rangle$, at least when $R$ increases to a certain finite value and in the thermodynamic limit. Note that, different from $|\Phi_b\rangle$ involving $C_2$-exchange PSG, a priori, there is no relation between the two colors' mean-field states in $|\Phi_a\rangle$. 

\begin{figure}
    \centering
    \includegraphics[width=0.7\linewidth]{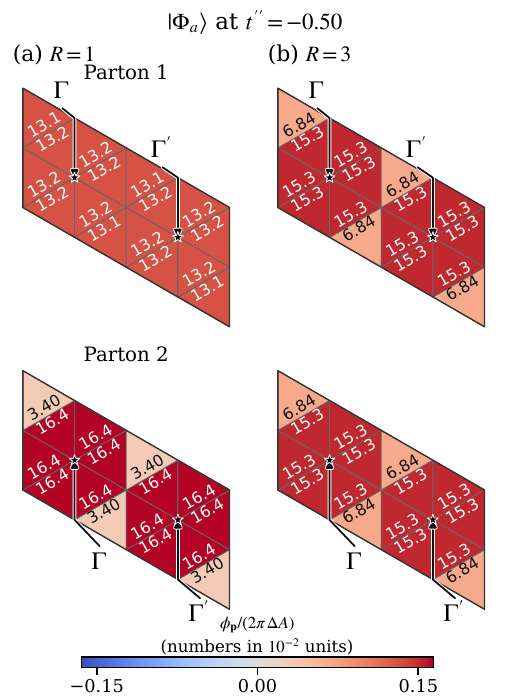}
    \caption{The Berry curvature densities for the two parton colors of the optimal (a) $R=1$ and (b) $R=3$ Hdet states of the FCI-a symmetry class (i.e., $|\Phi_a\rangle$) on the $N=4$ sample at $t''=-0.50$. The $R=3$ Berry curvature distribution is symmetric under parton color permutation $(1)\leftrightarrow(2)$.}
    \label{fig:Berry_FCIa}
\end{figure}

Although in the present simulation, we can only numerically access $R>1$ for the rather small $N=4$ sample, we can still test this argument by comparing $R=1$ and $R=3$. For this purpose, we compute the optimal parton mean-field states' Berry curvature distribution $\boldsymbol\phi^{(p)}_\mathbf p$ for the two colors $(1)$ and $(2)$ of $|\Phi_a\rangle$, as a function of $t''$. We show these distributions at $-t''=0.50$ in Fig.~\ref{fig:Berry_FCIa} as a demonstration. If there is an emergent color-permutation symmetry, the difference of the Berry curvature distribution between the two colors $\Delta \boldsymbol\phi _\mathbf p\equiv |\boldsymbol\phi_\mathbf p^{(1)}-\boldsymbol\phi_\mathbf p^{(2)}|$ should vanish.

\begin{figure}
    \centering
    \includegraphics[width=0.9\linewidth]{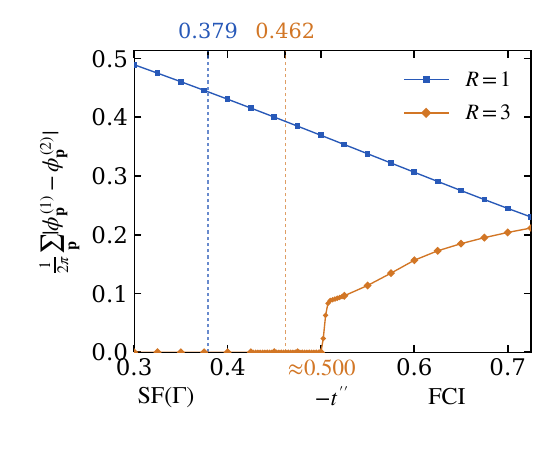}
    \caption{The difference in the Berry curvature distribution between the two parton colors for the optimal $R=1$ and $R=3$ Hdet states for the FCI-a symmetry class (i.e., $|\Phi_a\rangle$) on the $N=4$ sample, summed over all momentum space plaquettes $\mathbf p$. The $-t''$ values at the variational energy crossing points between $|\Phi_a\rangle$ and $|\Phi_{\Gamma}\rangle$ are marked on the figure top. For $R=3$  $|\Phi_a\rangle$ Hdet states, this difference vanishes near $-t''=0.500$.}
    \label{fig:berry_difference}
\end{figure}

The parton band inversion occurred from FCI to SF($\Gamma$) is reflected as a level crossing between $|\Phi_a\rangle$ and $|\Phi_\Gamma\rangle$, which happens at $t''= -0.379$ for $R=1$ and $t''= -0.462$ for $R=3$ on the $N=4$ sample, respectively (see Figs.~\ref{fig:N4R1} and~\ref{fig:N4R3}). For $R=1$, we observe that the $\Delta \boldsymbol\phi _\mathbf p$ remains finite throughout the FCI regime and increases when $t''$ is tuned towards the SF($\Gamma$) region. 

Strikingly, $\Delta \boldsymbol\phi _\mathbf p$
for $R=3$ is not only smaller than that for $R=1$, as one tunes $t''$ towards SF($\Gamma$), $\Delta \boldsymbol\phi _\mathbf p$ actually decreases and becomes zero up to the numerical optimization precision when $0.462<-t''<0.500$. We plot $\frac{1}{2\pi}\sum_\mathbf p\Delta\boldsymbol\phi_\mathbf p$ as a function of $t''$ in Fig.~\ref{fig:berry_difference}, which again highlights the importance of the new ingredient of the Hdet construction: $R>1$. If one uses only the traditional parton construction with $R=1$, the emergent parton color permutation symmetry will not be observed.

\section{Discussion and conclusion}
In this paper, we introduce the symmetry classes of the variational Hdet wavefunctions. Due to the parton Chern number conditions, these classes constrain the associated effective gauge theories and the long-range universal properties of these states. By constructing symmetry classes describing the FCI phase and the superfluid phases, we investigate the quantum phase diagram of a hard-core boson model. Compared with the traditional fermionic parton construction with $ R=1$ fusion channels, we show that the more general $R>1$ Hdet states significantly improve the microscopics and can quantitatively capture the phase diagram. 

As a byproduct, we argue that inside the FCI phase the Higgs transition $SU(2)_g\rightarrow U(1)_g$ is forbidden by the $C_6$ symmetry. In terms of the standard field-theoretical language, inside the FCI phase, the Lagrangian can be written as fermions coupled with the $SU(2)_g$ gauge field:
\begin{align}
L=&\sum_{I=1}^2\bar\psi_I(\sigma^{\mu} D_{\mu}+m+ g \Phi)\psi_I+L_{\rm gauge}[a]\notag\\
&+{\rm Tr}[D_{\mu}\Phi D_{\mu}\Phi]+r{\rm Tr}[\Phi^2]+u{\rm Tr}[\Phi^2]^2,
\end{align}
where $\sigma^\mu$ are the Dirac matrices (we choose $\sigma^0=\sigma_z,\sigma^1=\sigma_x,\sigma^2=\sigma_y$), $D_\mu$ are $SU(2)_g$-covariant derivatives, $I$ labels the two valleys, and $\Phi=\phi_x\tau_x+\phi_y\tau_y+\phi_z\tau_z$ is the $SU(2)_g$-adjoint Higgs field. The Higgs transition $SU(2)_g\rightarrow U(1)_g$ is described by $|\langle\Phi\rangle|\neq 0$. After fixing a uniform gauge, we may write $\langle\Phi\rangle=\phi_z\tau_z$ with $\phi_z\ge0$. If indeed $\phi_z>0$, the fermion masses for two colors, $\tau_z=\pm1$, are $m\pm g\phi_z$, and consequently the gap-closing transition involves only one color. As we have shown, \emph{in the same gauge}, the $C_2$ projective symmetry of the $C_6=1$ ground state (i.e., FCI-b) involves $\tau_x$, i.e., exchanging the two colors, enforcing $\phi_z$=0.

Precisely, after performing a long-wavelength expansion for the parton mean-field states and fixing a convenient gauge, following Eqs.~(\ref{eq:FCI-a-C_2-internal}) and~(\ref{eq:FCI-b-C_2-exchange}), one can show that the PSG element $U_{C_2}$ for the 180$^\circ$ rotation is realized in terms of $\psi$ as:
\begin{align}
&\text{FCI-a: } \psi_{I}(\mathbf r)\rightarrow(i\tau_z)(i\sigma_{z})\psi_{I}(-\mathbf r),\notag\\
&\text{FCI-b: } \psi_{I}(\mathbf r)\rightarrow(i\tau_x)(i\sigma_{z})\psi_{I}(-\mathbf r),
\end{align}
where the twofold ground states FCI-a ($C_6=e^{\frac{i2\pi}{3}}$) and FCI-b ($C_6=1$) feature different projective symmetry transformations and are related by a flat connection $\vec{\mathsf A}\cdot \tau_z$ (see Eq.~(\ref{eq:flat_connection})). Here $(i\sigma_{z})$ is the anticipated Dirac spinor transformation for $C_2$. Crucially, the color exchange $(i\tau_x)$ in the FCI-b transformation cannot be gauged away. This is because under $C_2$, $\vec{\mathsf A}$ changes sign and the color exchange is required to transform the state back. 

Consequently, at the critical point, the masses of the two colors need to vanish simultaneously. The natural critical theory for the putative SF($\Gamma$) to FCI transition should then be described by $N_f=2$ (from two valleys) $SU(2)_g$ QCD with $\langle\Phi\rangle=0$. Although in the present study, our argument for vanishing Higgs involves the $C_6$ PSG, there may be a deeper physical principle beyond the projective symmetry protecting the $SU(2)_g$ IGG, which we leave as a subject for future investigation. 

The present work is a static variational investigation that addresses only the ground states of the quantum system. The key advantage of the Hdet construction is that it allows for a mean-field-like description together with accurate energetics. For instance, in this work we have access to the parton mean-field states, allowing us to compute their Berry curvature distribution. What has not been achieved here is a full dynamical variational investigation capable of reliably computing the energetics of the anyonic excitations. For instance, it would be very nice to directly compute the parton Dirac spectrum near the SF($\Gamma$)-FCI transition, and observe the parton band inversion. Such a calculation must go beyond the variational Monte Carlo technique applied in the present work, since larger system sizes are required. In Ref.~\cite{Lin2026}, we outline such calculations based on the variational manifold path integral (VMPI) and the projective-expansion (PE). As a future project, we would like to apply these techniques to this hard-core boson model and perform a full dynamical calculation.

X.H.~and D.X.~were supported by the U.S. Department of Energy, Office of Basic Energy Sciences, Division of Materials Sciences and Engineering under Contract No. DE-AC02-76SF00515.

\begin{appendix}
\section{Derivation of the lattice PSG and boundary twists}
\label{app:psg_derivation}

We derive Eqs.~(\ref{eq:psg_group_identities}) and~(\ref{eq:psg-phase}) from
the spatial actions and Eq.~(\ref{eq:sg_identities}) in
Sec.~\ref{sec:psg}. As in the
main text, we consider $W_g=e^{i\phi_g(\mathbf r,\alpha)\tau_z}$ and
the $\pi$-flux condition Eq.~(\ref{eq:pi_flux_condition}). If needed, reflection can be imposed as an additional antiunitary symmetry.

\subsection{PSG convention and real-space gauge choice}

The PSG algebraic identities can initially carry arbitrary phases in the IGG. We use the global phase ambiguity in $W_{T_2}$ to fix Eq.~(\ref{eq:psg_rotation_1}). We then use the phase in $W_{T_1}$ to
fix Eq.~(\ref{eq:psg_rotation_2}), and finally the phase in
$W_{C_6}$ to fix Eq.~(\ref{eq:psg_rotation_6}). Each choice preserves the identities already fixed, and none changes the translation commutator Eq.~(\ref{eq:psg_translation}), which encodes the $\pi$ flux of our state. 
This specifies the PSG convention in Eq.~(\ref{eq:psg_group_identities}),
within which rotation compatibility determines the boundary twists.

We first consider color-$(1)$; the phases for color-$(2)$ have the
opposite sign. In $U_g=W_g g$, the phase is evaluated at the
destination site. A local gauge choice sets
$\phi_{T_1}=\varphi_1/N$ and
$\phi_{T_2}(0,y,w,\alpha)=\varphi_2/N$. The translation commutator
requires the $T_2$ phase to increase by $\pi$ when $x$ increases by
one. This gives the translation phases in Eq.~(\ref{eq:psg-phase}).

From Eqs.~(\ref{eq:psg_rotation_1}) and~(\ref{eq:psg_rotation_2}), we obtain
\begin{align}
&\phi_{C_6}(x+1,y,w,\alpha)-\phi_{C_6}(x,y,w,\alpha)
\notag\\
&\hspace{1cm}=\pi(x+y)+\frac{\varphi_2}{N},\notag\\
&\phi_{C_6}(x,y+1,w,\alpha)-\phi_{C_6}(x,y,w,\alpha)
\notag\\
&\hspace{1cm}=\pi x+\frac{\varphi_2-\varphi_1}{N}.
\label{eq:app_phase_recursions}
\end{align}
Summing these differences determines the coordinate-dependent terms
in Eq.~(\ref{eq:psg-phase}). Only
$\phi_{C_6}(0,0,w,\alpha)$ remains. Following a site through six
rotations and imposing $U_{C_6}^6=1$ gives
\begin{align}
&3\big[\phi_{C_6}(0,0,u,\alpha)+\phi_{C_6}(0,0,v,\alpha)\big]
\notag\\
&\hspace{1cm}-\pi+\frac{3\varphi_2}{N}=0\pmod{2\pi}.
\label{eq:app_rotation_closure}
\end{align}
A sublattice-dependent gauge transformation can be chosen to make the two constants
equal. Their common value can be chosen as
$-\pi/2-\varphi_2/(2N)+\pi l_\alpha/3$, reproducing
Eq.~(\ref{eq:psg-phase}). Shifting $l_\alpha$ by three is equivalent
to a relative sign between the sublattices, leaving the distinct
choices $l_\alpha=0,1,2$ specified in the main text. We use its
numerical choice $l_\alpha=0$ below.

\subsection{Boundary twists}

For even $N$, the translation commutator implies
\begin{align}
(U_{T_1}^{-1}U_{T_2})^N
=(-1)^{N(N-1)/2}U_{T_1}^{-N}U_{T_2}^N.
\label{eq:app_translation_power}
\end{align}
The sign counts the interchanges needed to put the translations in
the same order. Conjugating $U_{T_1}^N$ by $U_{C_6}$, using
Eq.~(\ref{eq:psg_group_identities}), requires
$e^{i\varphi_1}=e^{i\varphi_2}$. Conjugating $U_{T_2}^N$ and using
Eq.~(\ref{eq:app_translation_power}) fixes $e^{i\varphi_1}$ to
$(-1)^{N(N-1)/2}$. This proves Eq.~(\ref{eq:FCI_a_varphi}).

For the $C_3$ subgroup in Eq.~(\ref{eq:psg_subgroups}), the same
algebra gives
\begin{align}
U_{C_3}U_{T_1}U_{C_3}^{-1}&=U_{T_1}^{-1}U_{T_2},\notag\\
U_{C_3}U_{T_2}U_{C_3}^{-1}&=-U_{T_1}^{-1}.
\label{eq:app_C3_conjugation}
\end{align}
Taking the $N$th powers yields
\begin{align}
e^{i\varphi_1}&=(-1)^{N(N-1)/2}e^{-i\varphi_1+i\varphi_2},\notag\\
e^{i\varphi_2}&=e^{-i\varphi_1}.
\label{eq:app_C3_torus}
\end{align}
These are equivalent to Eq.~(\ref{eq:FCI_b_varphi}). Its solutions
are the FCI-a twists and the color-exchanged pair of FCI-b twists in
Eqs.~(\ref{eq:FCI_a_varphi}) and ~(\ref{eq:FCI_b_varphi_1}). In particular,
$\varphi_1=\varphi_2$ is not a consequence of $C_3$ symmetry alone.

For intra-color $C_2$, the algebra instead gives
$U_{C_2}U_{T_j}U_{C_2}^{-1}=-U_{T_j}^{-1}$ for $j=1,2$.
For even $N$, each twist is consequently zero or $\pi$. The SF($M$)
symmetry class uses the FCI-a twists within these choices, without
imposing intra-color $C_3$ or $C_6$.

For the FCI-b twists, spatial inversion reverses
$\boldsymbol\varphi$. The color exchange in
Eq.~(\ref{eq:FCI-b-C_2-exchange}) restores the original
opposite-color boundary conditions. Thus exchange-color $C_2$ is
compatible with these twists, whereas intra-color $C_2$ is not.

\section{Rotation indices and Chern-number constraints}
\label{app:psg_rep_derivation}

For a gapped filled band, we derive Eq.~(\ref{eq:C6_mod_12_constraint}) from the Berry flux in
the principal region in Fig.~\ref{fig:app_pr_bz}. After fixing the
gauge of the filled band basis, symmetry-related boundary lines
cancel. The remaining rotation indices give the projective
generalization of the Chern-number constraint \cite{Fang2012,JianQi2013}.

The $C_6$ indices in Eq.~(\ref{eq:C6_psg_indices}) are defined with $U_g$ satisfying
the following identities, as in
Eq.~(\ref{eq:psg_group_identities}):
\begin{align}
U_{C_6}^{6}&=1,\notag\\
U_{T_1}^{-1}U_{C_6}U_{T_1}U_{T_2}^{-1}U_{C_6}^{-1}&=1,\notag\\
U_{T_2}^{-1}U_{C_6}U_{T_1}U_{C_6}^{-1}&=1,
\label{eq:app_C6_index_identities}
\end{align}
Different $W_{C_6}$ gauge choices with $U_{C_6}^{6}\neq 1$ lead to different
rotation-index conventions.

\subsection{The principal region and the two pairs of lines}

\begin{figure}[tb]
\centering
\includegraphics[width=.57\columnwidth]{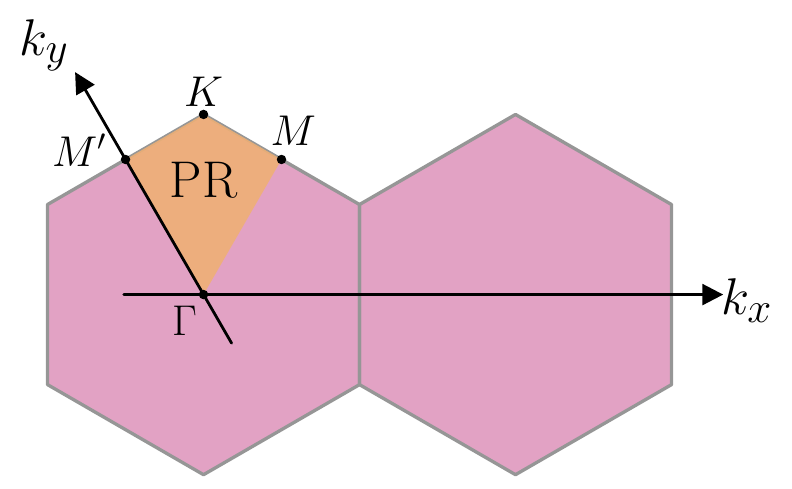}
\hfill
\includegraphics[width=.40\columnwidth]{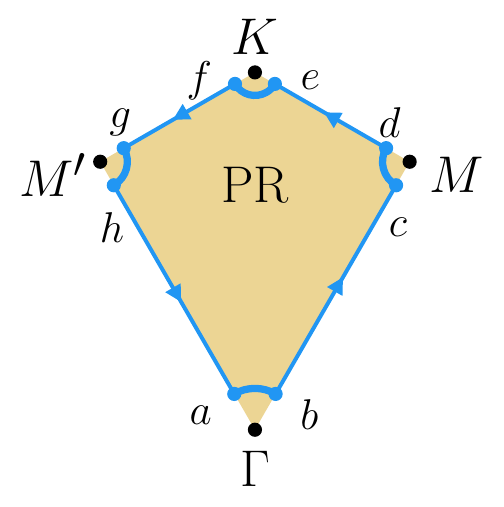}
\caption{Principal region for the $C_6$ Chern-number constraint.
The orange $\Gamma$--$M$--$K$--$M'$ region occupies one twelfth of
the magnetic BZ. $U_{C_6}$ relates $\Gamma$--$M$ to
$\Gamma$--$M'$, and $U_{T_2}U_{C_3}$ relates $K$--$M'$ to
$K$--$M$. The right diagram separates the lines from small arcs at
their endpoints. After phase fixing, the lines cancel and the
endpoint rotation matrix elements remain.}
\label{fig:app_pr_bz}
\end{figure}

The high-symmetry momenta are defined in
Sec.~\ref{sec:classify_symmetry_classes} and
Fig.~\ref{fig:parton_BZ}. For the contour in Fig.~\ref{fig:app_pr_bz},
use the periodically equivalent copies $M=(\pi,\pi)$,
$K=(5\pi/6,7\pi/6)$ and $M'=(\pi/2,\pi)$, with $\Gamma$
unchanged. These are the main-text points shifted by $(0,\pi)$,
not additional momentum points. The position-operator factor in
Eq.~(\ref{eq:berry_flux}) is retained across the BZ boundary. This region
has area $\pi^2/6$, one twelfth of the magnetic BZ. Its twelve
images have equal curvature by magnetic translation and $C_6$.

We consider the filled band of color-$(p)$. To show the cancellation
explicitly, we denote the phases of the two relevant rotation
matrix elements as
\begin{align}
e^{i\theta_6(\mathbf k)}
&=(-1)^{p-1}\sqrt2\,
\langle\phi^{(p)}_{R\mathbf k}|U_{C_6}|\phi^{(p)}_{\mathbf k}\rangle,
\notag\\
e^{i\theta_3(\mathbf k)}
&=\sqrt2\,
\langle\phi^{(p)}_{\mathbf q}|U_{T_2}U_{C_3}
|\phi^{(p)}_{\mathbf k}\rangle.
\label{eq:app_boundary_phases}
\end{align}
Here $\mathbf q=(2\pi-k_2,k_1-k_2+3\pi/2)$ is the output momentum
of $U_{T_2}U_{C_3}$ on this contour. Both phases have unit magnitude
and depend on the filled-band gauge at generic $\mathbf k$.

Equation~(\ref{eq:R_C6}) gives the first rotation matrix. The same
translation algebra gives the second one, in the same ordered
doublets $\{|\phi^{(p)}_{\mathbf k}\rangle,
U_{T_2}|\phi^{(p)}_{\mathbf k}\rangle\}$:
\begin{align}
\mathsf R^{(p)}_{T_2C_3}(\mathbf k)
=\frac{e^{i\theta_3(\mathbf k)}}{\sqrt2}
\begin{pmatrix}
1&e^{iq_1}\\
e^{i(q_1+k_1)}&e^{i(2q_2-k_1)}
\end{pmatrix}.
\label{eq:app_T2C3_matrix}
\end{align}
The matrix $\mathsf R^{(p)}_{T_2C_3}$ is defined by
Eq.~(\ref{eq:fixing_R_g}), applied to $U_{T_2}U_{C_3}$ with the
input and output doublets specified above.

\subsection{Cancellation of the open lines}

For neighboring momenta, use exactly the overlap with position
transport in Eq.~(\ref{eq:berry_flux}); choose the position origin at
the $C_6$ center. The two members of a magnetic doublet have opposite
$U_{T_1}$ eigenvalues, so their mixed overlaps vanish after momentum
transport. Their diagonal overlaps satisfy
\begin{align}
&\langle\phi^{(p)}_{\mathbf q'}|U_{T_2}^{\dagger}
e^{i(\mathbf q'-\mathbf q)\cdot\hat{\mathbf r}}U_{T_2}
|\phi^{(p)}_{\mathbf q}\rangle\notag\\
&\quad=e^{i(q'_2-q_2)}
\langle\phi^{(p)}_{\mathbf q'}|
e^{i(\mathbf q'-\mathbf q)\cdot\hat{\mathbf r}}
|\phi^{(p)}_{\mathbf q}\rangle.
\label{eq:app_partner_overlap}
\end{align}
The phase is just the one-cell translation in $U_{T_2}$.

First take $\mathbf k,\mathbf k'$ on $\Gamma$--$M$, and
$\mathbf q=R\mathbf k$, $\mathbf q'=R\mathbf k'$ on
$\Gamma$--$M'$. Substituting Eq.~(\ref{eq:R_C6}) and using
Eq.~(\ref{eq:app_partner_overlap}) gives
\begin{align}
&\langle\phi^{(p)}_{\mathbf k'}|
e^{i(\mathbf k'-\mathbf k)\cdot\hat{\mathbf r}}
|\phi^{(p)}_{\mathbf k}\rangle\notag\\
&\quad=e^{i[\theta_6(\mathbf k)-\theta_6(\mathbf k')]}
\langle\phi^{(p)}_{\mathbf q'}|
e^{i(\mathbf q'-\mathbf q)\cdot\hat{\mathbf r}}
|\phi^{(p)}_{\mathbf q}\rangle.
\label{eq:app_C6_link_cancellation}
\end{align}
Explicitly, the two diagonal terms contribute
$[1+e^{i(k_1-k'_1+q'_2-q_2)}]/2=1$, since $q_2=k_1$.
Multiplying normalized overlaps along $\Gamma\to M$ and
$M'\to\Gamma$ cancels their interior phases and leaves
$e^{i[\theta_6(\Gamma)-\theta_6(M)]}$. One can equivalently fix
$\theta_6=0$ on the open lines, with only endpoint phases remaining.

For the other pair, take $\mathbf k,\mathbf k'$ on $K$--$M'$ and
the output momenta of $U_{T_2}U_{C_3}$ specified above, on $K$--$M$.
The same calculation gives
\begin{align}
&\langle\phi^{(p)}_{\mathbf k'}|
e^{i(\mathbf k'-\mathbf k)\cdot\hat{\mathbf r}}
|\phi^{(p)}_{\mathbf k}\rangle\notag\\
&\quad=e^{-i(q'_2-q_2)}e^{i[\theta_3(\mathbf k)-\theta_3(\mathbf k')]}
\notag\\
&\qquad\times\langle\phi^{(p)}_{\mathbf q'}|
e^{i(\mathbf q'-\mathbf q)\cdot\hat{\mathbf r}}
|\phi^{(p)}_{\mathbf q}\rangle.
\label{eq:app_C3_link_cancellation}
\end{align}
The two diagonal terms again give one, using
$q_1+k_1=q_2+\pi/2$. Unlike the first pair, this operation contains
an explicit translation. Its Bloch phase gives the factor
$e^{-i(q'_2-q_2)}$, which must not be dropped. Along $K\to M'$,
the output $q_2$ changes by $-\pi/6$. Multiplying overlaps along
$M\to K$ and $K\to M'$ therefore leaves
$e^{i\pi/6}e^{i[\theta_3(K)-\theta_3(M')]}$.

The product around the counterclockwise boundary is the Berry flux
of this region, by Eq.~(\ref{eq:berry_flux}). Since it is one twelfth
of the magnetic BZ, the two cancellations yield
\begin{align}
e^{i\pi\mathcal C^{(p)}/6}
=&\ e^{i\pi/6}
e^{i[\theta_6(\Gamma)+\theta_3(K)]}\notag\\
&\times e^{-i[\theta_6(M)+\theta_3(M')]}.
\label{eq:app_endpoint_phases}
\end{align}
All open-line contributions cancel.

\subsection{The high-symmetry-point indices}

At the fixed corners, Eq.~(\ref{eq:C6_psg_indices}) identifies
$e^{i\theta_6(\Gamma)}=\alpha^{(p)}_\Gamma$ and
$e^{i\theta_3(K)}=\beta^{(p)}_K$. At $M,M'$, each phase depends
on the filled-band gauge. Their product follows by composing the
PSG operators. Equation~(\ref{eq:R_C6}) at $M$ and
Eq.~(\ref{eq:app_T2C3_matrix}) at $M'$ give the numerical product
\begin{align}
\frac{1}{2}
\begin{pmatrix}1&-1\\-i&-i\end{pmatrix}
\begin{pmatrix}1&-i\\-1&-i\end{pmatrix}
=\begin{pmatrix}1&0\\0&-1\end{pmatrix}.
\label{eq:app_M_matrix_product}
\end{align}
Its overall phase is $e^{i[\theta_6(M)+\theta_3(M')]}$.
The full operator product is
$(-1)^{p-1}U_{T_2}U_{C_3}U_{C_6}
=(-1)^{p-1}U_{T_2}U_{C_2}$. Its first diagonal element is precisely
$\gamma^{(p)}_M$ in Eq.~(\ref{eq:C6_psg_indices}). Hence
\begin{align}
e^{i[\theta_6(M)+\theta_3(M')]}=\gamma^{(p)}_M.
\label{eq:app_M_endpoint_phases}
\end{align}
The algebra also implies $(U_{T_2}U_{C_2})^2=-1$, so
$(\gamma^{(p)}_M)^{-1}=-\gamma^{(p)}_M$. Substituting into
Eq.~(\ref{eq:app_endpoint_phases}) leaves
$e^{-5\pi i/6}\alpha^{(p)}_\Gamma\beta^{(p)}_K\gamma^{(p)}_M$,
which proves Eq.~(\ref{eq:C6_mod_12_constraint}).

The allowed phases follow from the same matrices. At $\Gamma$, the
numerical matrix in Eq.~(\ref{eq:R_C6}), after removing
$\alpha^{(p)}_\Gamma$, has sixth power $iI_2$. At $K$, the
numerical matrix in Eq.~(\ref{eq:app_T2C3_matrix}), after removing
$\beta^{(p)}_K$, has cube $e^{i\pi/4}I_2$. Using $U_{C_6}^6=1$,
$(U_{T_2}U_{C_3})^3=-1$ and $(U_{T_2}U_{C_2})^2=-1$ gives
\begin{align}
(\alpha^{(p)}_\Gamma)^6&=-i,\notag\\
(\beta^{(p)}_K)^3&=e^{3\pi i/4},\notag\\
(\gamma^{(p)}_M)^2&=-1.
\label{eq:app_index_powers}
\end{align}
Their roots are exactly the $a^{(p)}_\Gamma,b^{(p)}_K,c^{(p)}_M$
parametrization in Eq.~(\ref{eq:C6_psg_indices}). Inserting those
phases into the endpoint product gives the integer relation in
Eq.~(\ref{eq:C6_mod_12_constraint}). Cubing it and using the second
line of Eq.~(\ref{eq:app_index_powers}) eliminates the $K$ index and
proves Eq.~(\ref{eq:C6_mod_4_constraint}).

\subsection{The \texorpdfstring{$C_3$}{C3} and \texorpdfstring{$C_2$}{C2} symmetry classes}

Using the same technique, one obtains
Eqs.~(\ref{eq:C3_mod_6_constraint}) and~(\ref{eq:C2_mod_4_constraint}).
The index definitions remain those in
Eqs.~(\ref{eq:C6_psg_indices}) and~(\ref{eq:other_psg_indices}). For the
PSG representative chosen in Eq.~(\ref{eq:psg_group_identities}),
the corresponding subgroup identities are
\begin{align}
U_{C_3}^{\,3}=1,\qquad &(U_{T_2}U_{C_3})^3=-1,\notag\\
U_{C_2}^{\,2}=1,\qquad &(U_{T_2}U_{C_2})^2=-1.
\label{eq:app_subgroup_power_identities}
\end{align}
For $C_3$, the first identity fixes the allowed $\Gamma$ phases,
and the second fixes the $K,K'$ phases. For intra-color $C_2$, the
first fixes the $\Gamma,M'$ phases and the second fixes the
$M,M''$ phases. The index assigned to a given filled state is
understood in the rotation-index convention specified by this $W_g$
gauge choice.

One further ingredient for $C_3$ is the odd Chern number of a gapped
filled magnetic band at the present half filling and $\pi$
flux \cite{Lu2020}. Combining this parity with the rotation
constraint gives the mod-six information in
Eq.~(\ref{eq:C3_mod_6_constraint}). The remaining boundary
cancellations require no separate technique.

\subsection{Indices accessible on the finite samples}

The boundary powers in Eq.~(\ref{eq:psg_group_identities}) and the
translation eigenvalues in Sec.~\ref{sec:classify_symmetry_classes}
give $k_j=[2\pi n_j-(-1)^{p-1}\varphi_j]/N$, modulo the magnetic
BZ. An index can be read directly only if its invariant momentum is
sampled. Table~\ref{tab:app_momentum_access} lists the possibilities
with the FCI-a twists.

\begin{table}[tb]
\caption{Accessible momenta with the FCI-a twists in
Eq.~(\ref{eq:FCI_a_varphi}). ``Yes'' means direct access to the
invariant-point index.}
\label{tab:app_momentum_access}
\centering
\setlength{\tabcolsep}{6pt}
\begin{tabular}{cccc}
\toprule
$N$ & $\Gamma$ & $\{K,K'\}$ & $\{M,M',M''\}$ \\
\midrule
4 & yes & no & yes\\
6 & yes & yes & no\\
8 & yes & no & yes\\
10 & yes & no & no\\
12 & yes & yes & yes\\
\bottomrule
\end{tabular}
\end{table}

The full $C_6$ mod-12 constraint requires both the $K$ and $M$
indices, so it is directly evaluable on this grid for $N$ a multiple
of 12. The mod-four reduction and the four-index intra-color-$C_2$
constraint are directly evaluable for $N$ a multiple of four. The
thermodynamic constraints refer to the specified PSG convention;
the boundary twists determine which indices are directly accessible
on a finite sample.

For the FCI-b twists, $\Gamma$ is absent on every finite even-$N$
grid. For $N=4,8,10$, one color samples $K$ and the other samples $K'$. However, on $N=6,12$ samples, neither $K$ nor $K'$ is sampled.

\section{Minimal parton mean-field states}
\label{app:min_mf}

As shown in Eq.~(\ref{eq:tensor_GL_inv}), the fusion tensor and the physical state are invariant under local $GL(R,\mathbb{C})$ transformations. Motivated by the notion of a \textit{minimal vector} in Kempf-Ness theory \cite{KempfNess1979}, we seek a representative that minimizes the norm of the \textit{unnormalized} parton mean-field state along its gauge orbit. 

For a pair of overlap matrices $\{ A^{(1)},A^{(2)}\}$, we denote the corresponding parton mean-field state by $|\Psi^{\text{MF}}(A^{(1)},A^{(2)})\rangle$. The inner product of two such states is
\begin{align}
	&\langle\Psi^{\text{MF}}(A^{(1)},A^{(2)})|
	\Psi^{\text{MF}}(B^{(1)},B^{(2)})\rangle\\
    \notag
    &\quad=\det(A^{(1)\dagger}B^{(1)})\det(A^{(2)\dagger}B^{(2)}).
\end{align}

By the polar decomposition, every $G \in GL(R,\mathbb{C})$ can be written as $G = UP$, where $U \in U(R)$ and $P$ is a positive definite Hermitian matrix. Writing $P = \exp(H)$, with $H^\dagger = H$, we obtain $G=U\exp(H)$. A local unitary transformation preserves the mean-field
norm, so changes in the norm arise from the Hermitian part of the
transformation.

For a local stretch $G_{\mathbf r}(s) = \exp(sH)$, the derivative of the squared norm at $s=0$ is
\begin{align}
	\left.\frac{d}{ds} \log \lVert\Psi^{\text{MF}}(s)\rVert^2 \right|_{s=0} &= 2\operatorname{Tr}[(\rho_{\mathbf r}^{(1)}-\rho_{\mathbf r}^{(2) T})H] 
\end{align}
The matrix
\begin{align}
    \mu_{\mathbf r}(A^{(1)},A^{(2)})
    \equiv
    \rho_{\mathbf r}^{(1)}-\rho_{\mathbf r}^{(2)T}
\end{align}
is the local moment map. Thus, a mean-field state that minimizes the norm
along its gauge orbit must satisfy $\mu_{\mathbf r}=0$ at every site
$\mathbf r$, which is precisely
Eq.~(\ref{eq:zero-moment-map condition}).
The zero-moment-map condition is preserved by local unitary
transformations, which belong to $U(R)$, the maximal compact subgroup
of $GL(R,\mathbb{C})$.

\section{Variational results for larger samples with $R=1$}
\label{app:larger_VMC}

\begin{figure*}[tb]
\centering
\includegraphics[width=0.9\textwidth]{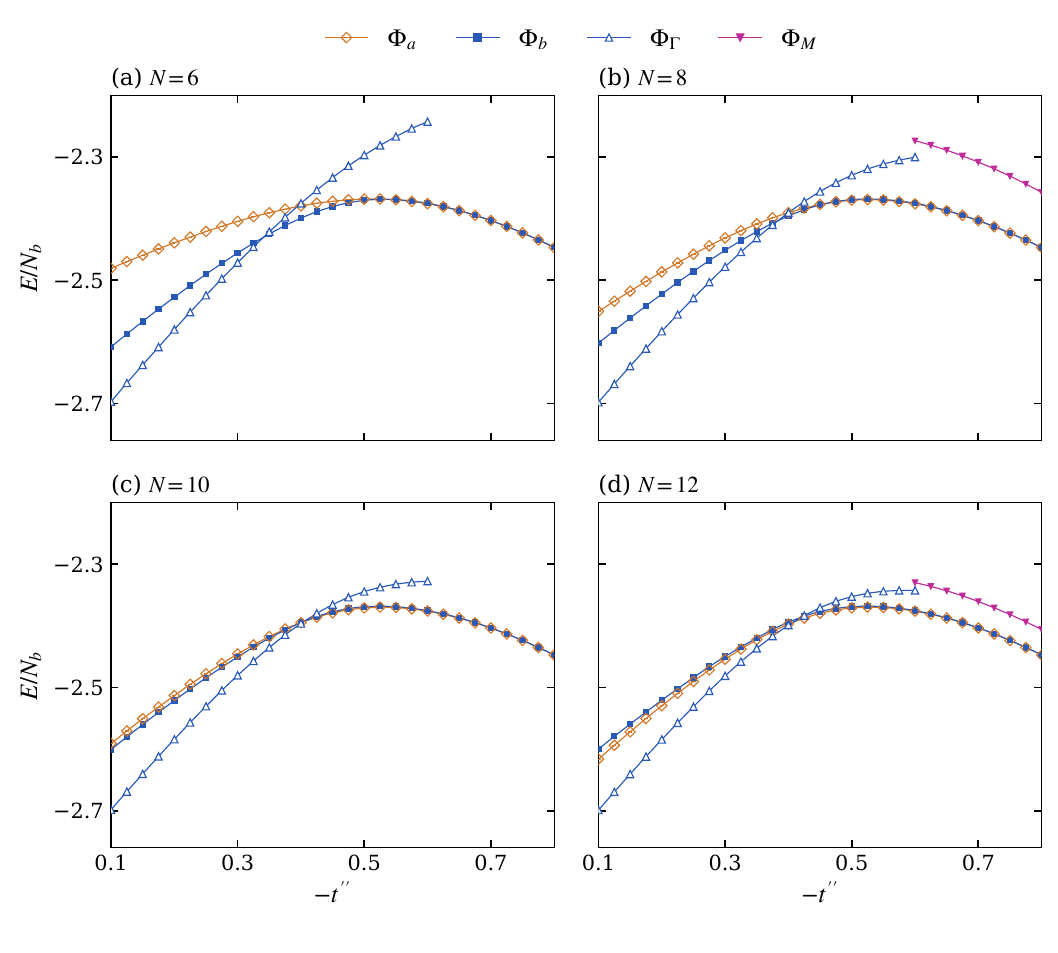}
\caption{Energy per boson of the symmetry classes in
Table~\ref{tab:classes}, for $R=1$ and (a) $N=6$, (b) $N=8$,
(c) $N=10$, (d) $N=12$.
The $|\Phi_\Gamma\rangle$ data cover $0.100\leq-t''\leq0.600$;
the $|\Phi_M\rangle$ data cover $0.600\leq-t''\leq0.800$ and
are shown only for $N=8,12$, where the required $C_2$ indices are
accessible. Monte Carlo error bars are smaller than the symbols.}
\label{fig:app_larger_energies}
\end{figure*}

We present the $R=1$ simulation results for $N=6,8,10,12$ in
Fig.~\ref{fig:app_larger_energies}, using the symmetry classes in
Table~\ref{tab:classes}. Note that $|\Phi_M\rangle$ is not shown
for $N=6,10$: the $M,M',M''$ points are not accessible on these
samples, as listed in Table~\ref{tab:app_momentum_access}.
Consequently, the $C_2$ indices cannot enforce the opposite parton
Chern numbers anticipated for the SF($M$) symmetry class.

\end{appendix}

% \clearpage

\bibliographystyle{apsrev4-2}
\bibliography{main_ref}

@article{Senthil2004,
  author = {Senthil, T. and Vishwanath, Ashvin and Balents, Leon and Sachdev, Subir and Fisher, Matthew P. A.},
  title = {Deconfined Quantum Critical Points},
  journal = {Science},
  volume = {303},
  pages = {1490--1494},
  year = {2004},
  doi = {10.1126/science.1091806},
  eprint = {cond-mat/0311326},
  archivePrefix = {arXiv}
}

@article{SenthilVojtaSachdev2004,
  author = {Senthil, T. and Vojta, Matthias and Sachdev, Subir},
  title = {Weak magnetism and non-{Fermi} liquids near heavy-fermion critical points},
  journal = {Phys. Rev. B},
  volume = {69},
  pages = {035111},
  year = {2004},
  doi = {10.1103/PhysRevB.69.035111},
  eprint = {cond-mat/0305193},
  archivePrefix = {arXiv}
}

@article{Senthil2008,
  author = {Senthil, T.},
  title = {Theory of a continuous {Mott} transition in two dimensions},
  journal = {Phys. Rev. B},
  volume = {78},
  pages = {045109},
  year = {2008},
  doi = {10.1103/PhysRevB.78.045109},
  eprint = {0804.1555},
  archivePrefix = {arXiv}
}

@article{Isakov2012,
  author = {Isakov, Sergei V. and Melko, Roger G. and Hastings, Matthew B.},
  title = {Universal Signatures of Fractionalized Quantum Critical Points},
  journal = {Science},
  volume = {335},
  pages = {193--195},
  year = {2012},
  doi = {10.1126/science.1212207},
  eprint = {1108.2055},
  archivePrefix = {arXiv}
}

@article{Nandkishore2012,
  author = {Nandkishore, Rahul and Metlitski, Max A. and Senthil, T.},
  title = {Orthogonal metals: The simplest non-{Fermi} liquids},
  journal = {Phys. Rev. B},
  volume = {86},
  pages = {045128},
  year = {2012},
  doi = {10.1103/PhysRevB.86.045128},
  eprint = {1201.5998},
  archivePrefix = {arXiv}
}

@article{SongZhang2023,
  author = {Song, Xue-Yang and Zhang, Ya-Hui},
  title = {Deconfined criticalities and dualities between chiral spin liquid, topological superconductor and charge density wave {Chern} insulator},
  journal = {SciPost Phys.},
  volume = {15},
  pages = {215},
  year = {2023},
  doi = {10.21468/SciPostPhys.15.5.215},
  eprint = {2206.08939},
  archivePrefix = {arXiv}
}

@article{Barkeshli2014,
  author = {Barkeshli, Maissam and McGreevy, John},
  title = {Continuous transition between fractional quantum {Hall} and superfluid states},
  journal = {Phys. Rev. B},
  volume = {89},
  pages = {235116},
  year = {2014},
  doi = {10.1103/PhysRevB.89.235116}
}

@article{BarkeshliYaoLaumann2015,
  author = {Barkeshli, Maissam and Yao, Norman Y. and Laumann, Christopher R.},
  title = {Continuous Preparation of a Fractional {Chern} Insulator},
  journal = {Phys. Rev. Lett.},
  volume = {115},
  pages = {026802},
  year = {2015},
  doi = {10.1103/PhysRevLett.115.026802}
}

@article{SongZhangSenthil2024,
  author = {Song, Xue-Yang and Zhang, Ya-Hui and Senthil, T.},
  title = {Phase transitions out of quantum {Hall} states in moir{\'e} materials},
  journal = {Phys. Rev. B},
  volume = {109},
  pages = {085143},
  year = {2024},
  doi = {10.1103/PhysRevB.109.085143},
  eprint = {2308.10903},
  archivePrefix = {arXiv}
}

@article{Lu2025,
  author = {Lu, Hongyu and Wu, Han-Qing and Chen, Bin-Bin and Meng, Zi Yang},
  title = {Continuous Transition between Bosonic Fractional {Chern} Insulator and Superfluid},
  journal = {Phys. Rev. Lett.},
  volume = {134},
  pages = {076601},
  year = {2025},
  doi = {10.1103/PhysRevLett.134.076601},
  eprint = {2405.18269},
  archivePrefix = {arXiv}
}

@misc{Wang2026,
  author = {Wang, Taige and Song, Xue-Yang and Zaletel, Michael P. and Senthil, T.},
  title = {Emergent {QED}$_3$ at the bosonic {Laughlin} state to superfluid transition},
  year = {2026},
  eprint = {2507.07611},
  archivePrefix = {arXiv},
  primaryClass = {cond-mat.str-el},
  doi = {10.48550/arXiv.2507.07611},
  url = {https://arxiv.org/abs/2507.07611}
}

@article{Barkeshli2012,
  author = {Barkeshli, Maissam and McGreevy, John},
  title = {Continuous transitions between composite {Fermi} liquid and {Landau Fermi} liquid: A route to fractionalized {Mott} insulators},
  journal = {Phys. Rev. B},
  volume = {86},
  pages = {075136},
  year = {2012},
  doi = {10.1103/PhysRevB.86.075136},
  eprint = {1206.6530},
  archivePrefix = {arXiv}
}

@article{Cai2023,
  author = {Cai, Jiaqi and Anderson, Eric and Wang, Chong and Zhang, Xiaowei and Liu, Xiaoyu and Holtzmann, William and Zhang, Yinong and Fan, Fengren and Taniguchi, Takashi and Watanabe, Kenji and Ran, Ying and Cao, Ting and Fu, Liang and Xiao, Di and Yao, Wang and Xu, Xiaodong},
  title = {Signatures of fractional quantum anomalous {Hall} states in twisted {MoTe}$_2$},
  journal = {Nature},
  volume = {622},
  pages = {63--68},
  year = {2023},
  doi = {10.1038/s41586-023-06289-w},
  eprint = {2304.08470},
  archivePrefix = {arXiv}
}

@article{Zeng2023,
  author = {Zeng, Yihang and Xia, Zhengchao and Kang, Kaifei and Zhu, Jiacheng and Kn{\"u}ppel, Patrick and Vaswani, Chirag and Watanabe, Kenji and Taniguchi, Takashi and Mak, Kin Fai and Shan, Jie},
  title = {Thermodynamic evidence of fractional {Chern} insulator in moir{\'e} {MoTe}$_2$},
  journal = {Nature},
  volume = {622},
  pages = {69--73},
  year = {2023},
  doi = {10.1038/s41586-023-06452-3}
}

@article{Park2023,
  author = {Park, Heonjoon and Cai, Jiaqi and Anderson, Eric and Zhang, Yinong and Zhu, Jiayi and Liu, Xiaoyu and Wang, Chong and Holtzmann, William and Hu, Chaowei and Liu, Zhaoyu and Taniguchi, Takashi and Watanabe, Kenji and Chu, Jiun-Haw and Cao, Ting and Fu, Liang and Yao, Wang and Chang, Cui-Zu and Cobden, David and Xiao, Di and Xu, Xiaodong},
  title = {Observation of fractionally quantized anomalous {Hall} effect},
  journal = {Nature},
  volume = {622},
  pages = {74--79},
  year = {2023},
  doi = {10.1038/s41586-023-06536-0},
  eprint = {2308.02657},
  archivePrefix = {arXiv}
}

@article{Lu2024,
  author = {Lu, Zhengguang and Han, Tonghang and Yao, Yuxuan and Reddy, Aidan P. and Yang, Jixiang and Seo, Junseok and Watanabe, Kenji and Taniguchi, Takashi and Fu, Liang and Ju, Long},
  title = {Fractional quantum anomalous {Hall} effect in multilayer graphene},
  journal = {Nature},
  volume = {626},
  pages = {759--764},
  year = {2024},
  doi = {10.1038/s41586-023-07010-7},
  eprint = {2309.17436},
  archivePrefix = {arXiv}
}

@article{Xu2023,
  author = {Xu, Fan and Sun, Zheng and Jia, Tongtong and Liu, Chang and Xu, Cheng and Li, Chushan and Gu, Yu and Watanabe, Kenji and Taniguchi, Takashi and Tong, Bingbing and Jia, Jinfeng and Shi, Zhiwen and Jiang, Shengwei and Zhang, Yang and Liu, Xiaoxue and Li, Tingxin},
  title = {Observation of Integer and Fractional Quantum Anomalous {Hall} Effects in Twisted Bilayer {MoTe}$_2$},
  journal = {Phys. Rev. X},
  volume = {13},
  pages = {031037},
  year = {2023},
  doi = {10.1103/PhysRevX.13.031037}
}

@article{Xie2025,
  author = {Xie, Jian and Huo, Zihao and Lu, Xin and Feng, Zuo and Zhang, Zaizhe and Wang, Wenxuan and Yang, Qiu and Watanabe, Kenji and Taniguchi, Takashi and Liu, Kaihui and Song, Zhida and Xie, X. C. and Liu, Jianpeng and Lu, Xiaobo},
  title = {Tunable fractional {Chern} insulators in rhombohedral graphene superlattices},
  journal = {Nat. Mater.},
  volume = {24},
  pages = {1042--1048},
  year = {2025},
  doi = {10.1038/s41563-025-02225-7},
  eprint = {2405.16944},
  archivePrefix = {arXiv}
}

@article{Haldane1988,
  author  = {Haldane, F. D. M.},
  title   = {Model for a Quantum Hall Effect without Landau Levels:
  Condensed-Matter Realization of the ``Parity Anomaly''},
  journal = {Phys. Rev. Lett.},
  volume  = {61},
  pages   = {2015--2018},
  year    = {1988},
  doi     = {10.1103/PhysRevLett.61.2015}
}

@article{Tang2011,
  author = {Tang, Evelyn and Mei, Jia-Wei and Wen, Xiao-Gang},
  title = {High-Temperature Fractional Quantum {Hall} States},
  journal = {Phys. Rev. Lett.},
  volume = {106},
  pages = {236802},
  year = {2011},
  doi = {10.1103/PhysRevLett.106.236802}
}

@article{Sun2011,
  author = {Sun, Kai and Gu, Zhengcheng and Katsura, Hosho and {Das Sarma}, S.},
  title = {Nearly Flatbands with Nontrivial Topology},
  journal = {Phys. Rev. Lett.},
  volume = {106},
  pages = {236803},
  year = {2011},
  doi = {10.1103/PhysRevLett.106.236803}
}

@article{Neupert2011,
  author = {Neupert, Titus and Santos, Luiz and Chamon, Claudio and Mudry, Christopher},
  title = {Fractional Quantum {Hall} States at Zero Magnetic Field},
  journal = {Phys. Rev. Lett.},
  volume = {106},
  pages = {236804},
  year = {2011},
  doi = {10.1103/PhysRevLett.106.236804}
}

@article{Sheng2011,
  author = {Sheng, D. N. and Gu, Zheng-Cheng and Sun, Kai and Sheng, L.},
  title = {Fractional quantum {Hall} effect in the absence of {Landau} levels},
  journal = {Nat. Commun.},
  volume = {2},
  pages = {389},
  year = {2011},
  doi = {10.1038/ncomms1380},
  eprint = {1102.2658},
  archivePrefix = {arXiv}
}

@article{RegnaultBernevig2011,
  author = {Regnault, N. and Bernevig, B. Andrei},
  title = {Fractional {Chern} Insulator},
  journal = {Phys. Rev. X},
  volume = {1},
  pages = {021014},
  year = {2011},
  doi = {10.1103/PhysRevX.1.021014}
}

@article{Xiao2011,
  author = {Xiao, Di and Zhu, Wenguang and Ran, Ying and Nagaosa, Naoto and Okamoto, Satoshi},
  title = {Interface engineering of quantum {Hall} effects in digital transition metal oxide heterostructures},
  journal = {Nat. Commun.},
  volume = {2},
  pages = {596},
  year = {2011},
  doi = {10.1038/ncomms1602},
  eprint = {1106.4296},
  archivePrefix = {arXiv}
}

@article{He2017,
  author  = {He, Yin-Chen and Zaletel, Michael P. and
             Oshikawa, Masaki and Pollmann, Frank},
  title   = {Signatures of Dirac Cones in a {DMRG} Study of the
             Kagome Heisenberg Model},
  journal = {Phys. Rev. X},
  volume  = {7},
  pages   = {031020},
  year    = {2017},
  doi     = {10.1103/PhysRevX.7.031020}
}

@article{Essin2014,
  author = {Essin, Andrew M. and Hermele, Michael},
  title = {Spectroscopic signatures of crystal momentum fractionalization},
  journal = {Phys. Rev. B},
  volume = {90},
  pages = {121102(R)},
  year = {2014},
  doi = {10.1103/PhysRevB.90.121102},
  eprint = {1401.1846},
  archivePrefix = {arXiv}
}

@article{Baskaran1987,
  author  = {Baskaran, G. and Zou, Z. and Anderson, P. W.},
  title   = {The resonating valence bond state and high-$T_c$
             superconductivity: A mean field theory},
  journal = {Solid State Commun.},
  volume  = {63},
  pages   = {973--976},
  year    = {1987},
  doi     = {10.1016/0038-1098(87)90642-9}
}

@article{AffleckMarston1988,
  author  = {Affleck, Ian and Marston, J. Brad},
  title   = {Large-$n$ limit of the {Heisenberg}-{Hubbard} model:
             Implications for high-$T_c$ superconductors},
  journal = {Phys. Rev. B},
  volume  = {37},
  pages   = {3774--3777},
  year    = {1988},
  doi     = {10.1103/PhysRevB.37.3774}
}

@article{Jain1989,
  author = {Jain, J. K.},
  title = {Incompressible quantum {Hall} states},
  journal = {Phys. Rev. B},
  volume = {40},
  pages = {8079--8082},
  year = {1989},
  doi = {10.1103/PhysRevB.40.8079}
}

@article{Wen1991,
  author = {Wen, X. G.},
  title = {Mean-field theory of spin-liquid states with finite energy gap and topological orders},
  journal = {Phys. Rev. B},
  volume = {44},
  pages = {2664--2672},
  year = {1991},
  doi = {10.1103/PhysRevB.44.2664}
}

@article{LeeNagaosaWen2006,
  author  = {Lee, Patrick A. and Nagaosa, Naoto and Wen, Xiao-Gang},
  title   = {Doping a {Mott} insulator: Physics of
             high-temperature superconductivity},
  journal = {Rev. Mod. Phys.},
  volume  = {78},
  pages   = {17--85},
  year    = {2006},
  doi     = {10.1103/RevModPhys.78.17}
}

@article{Ran2007,
  author = {Ran, Ying and Hermele, Michael and Lee, Patrick A. and Wen, Xiao-Gang},
  title = {Projected-Wave-Function Study of the Spin-$1/2$ {Heisenberg} Model on the Kagom{\'e} Lattice},
  journal = {Phys. Rev. Lett.},
  volume = {98},
  pages = {117205},
  year = {2007},
  doi = {10.1103/PhysRevLett.98.117205}
}

@article{McGreevy2012,
  author = {McGreevy, John and Swingle, Brian and Tran, Ky-Anh},
  title = {Wave functions for fractional {Chern} insulators},
  journal = {Phys. Rev. B},
  volume = {85},
  pages = {125105},
  year = {2012},
  doi = {10.1103/PhysRevB.85.125105},
  eprint = {1109.1569},
  archivePrefix = {arXiv}
}

@article{Lotric2026,
  author = {Lotri{\v c}, Tev{\v z} and Simon, Steven H.},
  title = {Paired Parton Trial States for the Superfluid-Fractional {Chern} Insulator Transition},
  journal = {Phys. Rev. Lett.},
  volume = {136},
  pages = {096601},
  year = {2026},
  doi = {10.1103/l9rp-y68y}
}

@article{Hu2024,
  author = {Hu, Xiaodong and Xiao, Di and Ran, Ying},
  title = {Hyperdeterminants and composite fermion states in fractional {Chern} insulators},
  journal = {Phys. Rev. B},
  volume = {109},
  pages = {245125},
  year = {2024},
  doi = {10.1103/PhysRevB.109.245125},
  eprint = {2312.00636},
  archivePrefix = {arXiv}
}

@misc{Lin2026,
  author = {Lin, Guan-Lin and Xiao, Di and Ran, Ying},
  title = {Hyperdeterminant wavefunctions},
  year = {2026},
  eprint = {2607.23392},
  archivePrefix = {arXiv},
  primaryClass = {cond-mat.str-el},
  doi = {10.48550/arXiv.2607.23392},
  url = {https://arxiv.org/abs/2607.23392}
}

@article{Wen2002,
  author = {Wen, Xiao-Gang},
  title = {Quantum orders and symmetric spin liquids},
  journal = {Phys. Rev. B},
  volume = {65},
  pages = {165113},
  year = {2002},
  doi = {10.1103/PhysRevB.65.165113},
  eprint = {cond-mat/0107071},
  archivePrefix = {arXiv}
}

@article{FisherLee1989,
  author = {Fisher, Matthew P. A. and Lee, D. H.},
  title = {Correspondence between two-dimensional bosons and a bulk superconductor in a magnetic field},
  journal = {Phys. Rev. B},
  volume = {39},
  pages = {2756--2759},
  year = {1989},
  doi = {10.1103/PhysRevB.39.2756}
}

@article{Wang2011,
  author = {Wang, Yi-Fei and Gu, Zheng-Cheng and Gong, Chang-De and Sheng, D. N.},
  title = {Fractional Quantum {Hall} Effect of Hard-Core Bosons in Topological Flat Bands},
  journal = {Phys. Rev. Lett.},
  volume = {107},
  pages = {146803},
  year = {2011},
  doi = {10.1103/PhysRevLett.107.146803},
  eprint = {1103.1686},
  archivePrefix = {arXiv}
}

@article{White1992,
  author = {White, Steven R.},
  title = {Density matrix formulation for quantum renormalization groups},
  journal = {Phys. Rev. Lett.},
  volume = {69},
  pages = {2863--2866},
  year = {1992},
  doi = {10.1103/PhysRevLett.69.2863}
}

@misc{Hu2026,
  author = {Hu, Xiaodong and Lin, Guan-Lin and Ran, Ying and Xiao, Di},
  title = {{HyperDet} Wavefunction: A Phase-Agnostic Ansatz for Strongly Correlated Systems},
  year = {2026},
  eprint = {2609.04146},
  archivePrefix = {arXiv},
  primaryClass = {cond-mat.str-el},
  doi = {10.48550/arXiv.2609.04146},
  url = {https://arxiv.org/abs/2609.04146}
}

@article{Zhang2012,
  author = {Zhang, Yi and Grover, Tarun and Turner, Ari and Oshikawa, Masaki and Vishwanath, Ashvin},
  title = {Quasiparticle statistics and braiding from ground-state entanglement},
  journal = {Phys. Rev. B},
  volume = {85},
  pages = {235151},
  year = {2012},
  doi = {10.1103/PhysRevB.85.235151}
}

@article{JianQi2013,
  author  = {Jian, Chao-Ming and Qi, Xiao-Liang},
  title   = {Crystal-symmetry preserving {Wannier} states for
             fractional {Chern} insulators},
  journal = {Phys. Rev. B},
  volume  = {88},
  pages   = {165134},
  year    = {2013},
  doi     = {10.1103/PhysRevB.88.165134}
}

@article{Fang2012,
  author = {Fang, Chen and Gilbert, Matthew J. and Bernevig, B. Andrei},
  title = {Bulk topological invariants in noninteracting point group symmetric insulators},
  journal = {Phys. Rev. B},
  volume = {86},
  pages = {115112},
  year = {2012},
  doi = {10.1103/PhysRevB.86.115112},
  eprint = {1207.5767},
  archivePrefix = {arXiv}
}

@article{FangCano2023,
  author  = {Fang, Yuan and Cano, Jennifer},
  title   = {Symmetry indicators in commensurate magnetic flux},
  journal = {Phys. Rev. B},
  volume  = {107},
  pages   = {245108},
  year    = {2023},
  doi     = {10.1103/PhysRevB.107.245108}
}

@incollection{KempfNess1979,
  author = {Kempf, George and Ness, Linda},
  title = {The length of vectors in representation spaces},
  booktitle = {Algebraic Geometry},
  editor = {L{\o}nsted, Knud},
  series = {Lecture Notes in Mathematics},
  volume = {732},
  pages = {233--243},
  publisher = {Springer},
  address = {Berlin, Heidelberg},
  year = {1979},
  doi = {10.1007/BFb0066647}
}

@book{BeccaSorella2017,
  author = {Becca, Federico and Sorella, Sandro},
  title = {Quantum Monte Carlo Approaches for Correlated Systems},
  publisher = {Cambridge University Press},
  year = {2017}
}

@article{Sorella2001,
  author = {Sorella, Sandro},
  title = {Generalized {Lanczos} algorithm for variational quantum {Monte Carlo}},
  journal = {Phys. Rev. B},
  volume = {64},
  pages = {024512},
  year = {2001},
  doi = {10.1103/PhysRevB.64.024512},
  eprint = {cond-mat/0009149},
  archivePrefix = {arXiv}
}

@article{Sorella2005,
  author = {Sorella, Sandro},
  title = {Wave function optimization in the variational {Monte Carlo} method},
  journal = {Phys. Rev. B},
  volume = {71},
  pages = {241103(R)},
  year = {2005},
  doi = {10.1103/PhysRevB.71.241103}
}

@article{You2007,
  author = {You, Wen-Long and Li, Ying-Wai and Gu, Shi-Jian},
  title = {Fidelity, dynamic structure factor, and susceptibility in critical phenomena},
  journal = {Phys. Rev. E},
  volume = {76},
  pages = {022101},
  year = {2007},
  doi = {10.1103/PhysRevE.76.022101},
  eprint = {quant-ph/0701077},
  archivePrefix = {arXiv}
}

@article{Wahl2013,
  author = {Wahl, T. B. and Tu, H.-H. and Schuch, N. and Cirac, J. I.},
  title = {Projected Entangled-Pair States Can Describe Chiral Topological States},
  journal = {Phys. Rev. Lett.},
  volume = {111},
  pages = {236805},
  year = {2013},
  doi = {10.1103/PhysRevLett.111.236805}
}

@article{DubailRead2015,
  author = {Dubail, J. and Read, N.},
  title = {Tensor network trial states for chiral topological phases in two dimensions and a no-go theorem in any dimension},
  journal = {Phys. Rev. B},
  volume = {92},
  pages = {205307},
  year = {2015},
  doi = {10.1103/PhysRevB.92.205307},
  eprint = {1307.7726},
  archivePrefix = {arXiv}
}

@article{Lu2020,
  author = {Lu, Yuan-Ming and Ran, Ying and Oshikawa, Masaki},
  title = {Filling-enforced constraint on the quantized {Hall} conductivity on a periodic lattice},
  journal = {Ann. Phys.},
  volume = {413},
  pages = {168060},
  year = {2020},
  doi = {10.1016/j.aop.2019.168060},
  eprint = {1705.09298},
  archivePrefix = {arXiv}
}

\end{document}